\documentclass{article}

\usepackage[noadjust]{cite}
\usepackage{amsmath,amssymb,amsfonts,amsthm,bbm}
\usepackage{algorithmicx}
\usepackage{textcomp}
\usepackage{xcolor}
\def\BibTeX{{\rm B\kern-.05em{\sc i\kern-.025em b}\kern-.08em
    T\kern-.1667em\lower.7ex\hbox{E}\kern-.125emX}}

\usepackage[ansinew]{inputenc}          
\usepackage{psfrag}                     
\usepackage[dvips]{graphicx,gincltex,epsfig}            
\usepackage[all]{xy}
\usepackage{a4wide}
\usepackage[dvips]{epsfig}
\usepackage{enumerate}
\usepackage{color}
\usepackage[color=green!30]{todonotes}
\usepackage{tikz}
\usepackage{nccmath}
\usepackage{hhline} 
\makeatletter
\newcommand\HUGE{\@setfontsize\Huge{34}{60}}
\makeatother   
\usepackage{enumitem, hyperref}
\makeatletter
\def\namedlabel#1#2{\begingroup
    #2%
    \def\@currentlabel{#2}%
    \phantomsection\label{#1}\endgroup
}
\makeatother

\usepackage{algorithm,algpseudocode}

\theoremstyle{plain}
\newtheorem{theorem}{Theorem}

\newtheorem{lemma}[theorem]{Lemma}

\newtheorem{proposition}[theorem]{Proposition}
\theoremstyle{definition}

\newtheorem{remark}[theorem]{Remark}

\newcommand{\Z}{\mathbb{Z}}

\newcommand{\F}{\mathbb{Z}}

\newcommand{\nolla}{\mathbf{0}}
\newcommand{\bu}{\mathbf{u}}

\newcommand{\bw}{\mathbf{w}}

\newcommand{\s}{\mathbf{s}}
\newcommand{\bc}{\mathbf{c}}

\newcommand{\supp}{\textrm{supp}}
\newcommand{\vsupp}{\textrm{vsupp}}

\newcommand{\bx}{\mathbf{x}}

\newcommand{\bz}{\mathbf{z}}

\newcommand{\y}{\mathbf{y}}
\newcommand{\by}{\mathbf{y}}

\newcommand{\LL}{\mathcal{L}}
\newcommand{\be}{\mathbf{e}}

\newcommand{\com}{\mathcal{COM}}
\newcommand{\eras}{\ast}

\title{Levenshtein's Sequence Reconstruction Problem and Results for  Larger Alphabet Sizes
	\thanks{The authors were funded in part by the Academy of Finland  grants 338797 and 358718. 	Some of the results in this article were presented without proofs in ISIT2023 \cite{junnila2023levenshtein}.}
}

\author{\textbf{Ville Junnila}, \textbf{Tero Laihonen} and \textbf{Tuomo Lehtil{\"a}\thanks{Corresponding author.}}\\
	Department of Mathematics and Statistics\\
	University of Turku, FI-20014 Turku, Finland\\
	viljun@utu.fi, terolai@utu.fi and tualeh@utu.fi}
\begin{document}

\maketitle


\begin{abstract}
        The problem of storing large amounts of information safely for a long period of time has become essential. One of the most promising new data storage mediums are the polymer-based data storage systems, like the DNA-storage system. These storage systems are highly durable and they consume very little energy to store the data. When information is retrieved from a storage, however, several different types of errors may occur in the process. It is known that the Levenshtein's sequence reconstruction framework is well-suited to overcome such errors and to retrieve the original information. Many of the previous results regarding Levenshtein's sequence reconstruction method are so far given only for the binary alphabet. However, larger alphabets are natural for the polymer-based data storage. For example, the quaternary alphabet is suitable for DNA-storage due to the four amino-acids in DNA. The results for larger alphabets often require, as we will see in this work, different and more complicated techniques compared to the binary case. Moreover, we show that an increase in the alphabet size makes some error types behave rather surprisingly.
\end{abstract}


\section{Introduction}\label{SecIntro}

The \textit{Levenshtein's sequence reconstruction problem}  was introduced in \cite{Levenshtein}. The setup of the problem is as follows. We have a subset $C$ of words of length $n$ where the alphabet has size $q$. A word $\bx$ of $C$ is transmitted through $N$ channels and in each of them some errors occur, for example, substitution, deletion or insertion errors (these error types are defined below) to the word $\bx$. With the aid of the (erroneous) output words $\by_1,\by_2,\dots,\by_N$ from the channels, we try to determine, i.e., reconstruct, the transmitted word $\bx$. Sometimes, when we cannot determine $\bx$ uniquely, we wish to have as small list as possible of candidates for $\bx$, which is naturally included in the list. Let us look at a small example. Suppose that $C$ consists of all the binary (i.e., $q=2$) words of length 3 and we transmit $\bx=000$ through two channels (i.e., $N=2$). In each channel it is assumed that at most one substitution error occurs, that is, in one coordinate the symbol 0 can change to 1 or vice versa. Suppose we  obtain the output words $\by_1=100$ and $\by_2= 010$ from the channels and we denote the set of output words by $Y=\{\by_1,\by_2\}$. Now we consider all the possible words in $C$,  denote this set by $T(Y)$, that could have been sent when we receive the above set $Y$, keeping in mind, that at most one substitution error can occur in the channels (the set $T(Y)$ is given more precisely in \eqref{TY} below). The words that could have been transmitted are $000$ and $110$. Hence, $T(Y) = \{000, 110\}$ and the length of the list is equal to two. Notice that $\bx\in T(Y).$ If we have more channels, say $N=3$, and, for example, the set of output words is $Y=\{100, 010,001\}$, then we can determine the transmitted word unambiguously since now $T(Y)=\{\bx\}$. In fact, it follows by Theorem~\ref{LevL=1} that if we send any word $\bx$ of $C$, then with the aid of three distinct output words, we can always determine $\bx$ uniquely.

 The topic of Levenshtein's sequence reconstruction problem has been widely studied during recent years \cite{yaakobi2018uncertainty, horovitz2018reconstruction, gabrys2018sequence, Maria_Abu-Sini, abu2023intersection, Uusi_Maria_Abu-Sini, junnila2022levenshtein, junnila2020levenshtein, goyal2022sequence, sabary2024survey}. Levenshtein's original motivation came from molecular biology and chemistry, where  adding redundancy was not feasible. Levenshtein's problem has recently become again significant due to advanced memory storage technologies such as associative memories \cite{yaakobi2018uncertainty}, racetrack memories \cite{chee2018reconstruction} and, especially, polymer-based memories like DNA-memories \cite{Uusi_Maria_Abu-Sini}, where the information is stored in the DNA molecules. The DNA-storage process consists of three steps, namely, \emph{synthesizing} (writing the information into DNA), \emph{storing} and \emph{sequencing} (retrieving the information). The first step, synthesis, produces artificial DNA molecules called \emph{strands} to encode the user's information units. Due to technical reasons, this phase produces several noisy strands of the encoded data. In the second phase, the DNA strands are stored  in an unordered manner in a storage container, where some molecules might be lost due to decay. The final sequencing step involves obtaining numerous erroneous copies of a stored synthesized strand. From these erroneous copies we should determine the original information in the strand. Synthesizing and sequencing DNA cause substitution, insertion and deletion errors to the information (these error types are discussed in more details below). This paper concentrates on the last phase of the process, that is, the information retrieval part. As multiple erroneous copies (corresponding to the output words in $Y$ discussed above) of the stored information unit $\bx$ are obtained,  the Levenshtein's model is very suitable for this problem where we wish to determine (reconstruct) $\bx$ using $Y$. Another interesting property which we obtain from DNA-applications and polymer-based memory systems in general, is the emphasis on $q$-ary $(q>2)$ information over the binary (see \cite{bornholt2016dna, church2012next, grass2015robust, yazdi2015dna} for more information about DNA-memories).

Let us next consider some notations. We will denote the set $\{1,2,\dots,n\}$ by $[1,n]$ and by $\F_q^n$ the set of $q$-ary words of length $n$ (over the alphabet $\F_q$). The set $\F_q^n$ is often called the \textit{$q$-ary $n$-dimensional Hamming space}. The  \emph{support} of a word $\bx=x_1\cdots x_n\in \F_q^n$ is defined by $\supp(\bx)=\{i\mid x_i\neq 0\}$, the \textit{weight} of $\bx$ by $w(\bx)=|\supp(\bx)|$ and the \textit{Hamming distance} between $\bx$ and $\by$ by $d(\bx,\by)=w(\bx-\by)$. For the \textit{Hamming balls} we use notation $B_t(\bx)=\{\by\in\F_q^n\mid d(\bx,\by)\leq t\}$ and $|B_t(\bx)|=V_q(n,t)=\sum_{i=0}^t(q-1)^i\binom{n}{i}$. A \textit{code} $C$ is a nonempty subset of $\F^n_q$ and it has \textit{minimum distance} $$d_{\min}(C)=\min_{\bc_1,\bc_2\in C,\bc_1\neq\bc_2}d(\bc_1,\bc_2).$$ Furthermore, $C$ is \textit{$e$-error-correcting} if $d_{\min}\geq2e+1$. Finally, we denote the \textit{zero-word} $00\cdots0$ with $\nolla$.
The   \textit{value support} of word $\bx$ is defined by $\vsupp(\bx)=\{(i,x_i)\mid x_i\neq 0\}$. In particular, we have
\begin{equation}\label{vsupp} d(\bx,\by)=w(\bx)+w(\by)-|\supp(\bx)\cap\supp(\by)|-|\vsupp(\bx)\cap\vsupp(\by)|.
\end{equation}
This implies that $d(\bx,\by)\geq w(\bx)+w(\by)-2|\supp(\bx)\cap\supp(\by)|$.

 Let us discuss the following three types of errors that are particularly relevant for DNA-memories \cite{heckel2019characterization}.
 In a \textit{substitution error} a symbol in some coordinate position is substituted with another symbol of the alphabet, in an \textit{insertion error} a new symbol of the alphabet is inserted  to (any position) in the original word leading to a word of length $n+1$ and in a \textit{deletion error} a symbol is deleted from (any position of) the original word leading to a word of length $n-1$. For example, if $\bx=014\in \F_5^3$, then the words $0104$ and $3014$ are obtained using one insertion error and the words $14$ and $04$ are obtained using one deletion to $\bx$. 
 In this paper, we also consider another usual error type discussed in coding theory, namely, the \textit{erasure errors}. 
  In an \textit{erasure error}, we replace  $x_i$, the $i$th symbol of $\bx$, with the  symbol $\eras$ representing a coordinate in an output word where we cannot read the symbol. For example, if $\bx=014$, then the words $\eras 14$ and $0\!\eras\!4$ are obtained by one erasure error. 
  
Depending on the error type, we may need also other types of balls than just the typical Hamming balls which are suitable for substitution errors. By $B^{e,s}_{t_e,t_s}(\bx)$ we denote the ball containing all words which can be obtained from $\bx$ with at most $t_e$ erasures and at most $t_s$ substitutions. By $B^d_{t_d}(\bx)$ we denote the set (ball) of words which can be obtained with at most $t_d$ deletions from $\bx$ and by $B^e_{t_e}(\bx)$ we denote the set of words which can be obtained with at most $t_e$ erasures from $\bx$.

\begin{figure}[tp]
	\centering
	\includegraphics[scale=0.30,trim=1500 271 1500 68,clip]{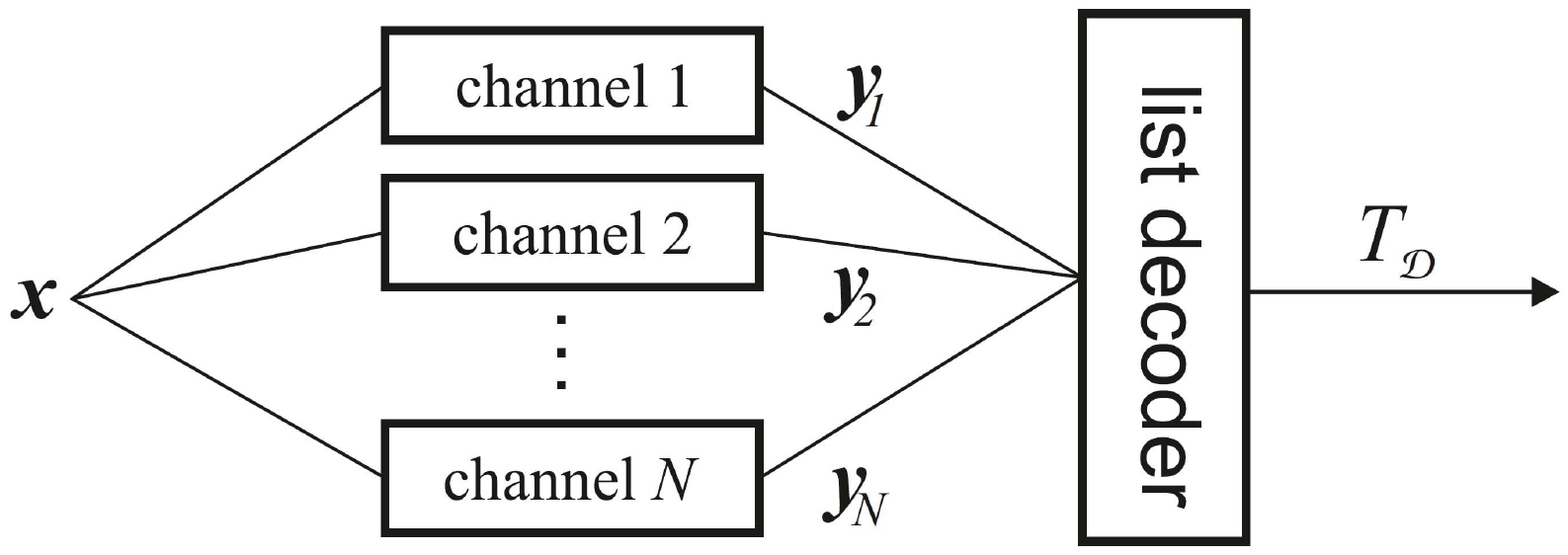}
	\centering\caption{The Levenshtein's sequence reconstruction.} \label{LevenshteinFig}
\end{figure}

Now, we  define the sequence reconstruction problem more rigorously. For the rest of the paper, we assume the following: $C\subseteq \F^n_q$ is a  code, a \textit{transmitted word} $\bx\in C$ is sent through $N$ channels in which substitution errors (or, depending on the case, insertion, deletion or erasure errors) may occur. Also we assume that the number of each type of error is limited by some constant $t$. Furthermore, we obtain a set $Y = \{\by_1, \by_2, \ldots, \by_N \}$ of output words from the $N$ channels.
 Depending on the situation, we may assume that each channel gives a different output word, that is, $Y$ is a \emph{set} of output words,  or  some output words can be the same, that is, $Y$ is a \emph{multiset} (in this paper, apart from Section~\ref{SecLargeq}, we consider $Y$ as a set). Based on the (multi)set of output words $Y$, we try to deduce the transmitted word $\bx$. However, sometimes we cannot do it and instead we have a list of possible transmitted words $T(Y)$ such that $\bx\in T(Y)$. The maximum size of this list is denoted by $\LL$. Sometimes we use the notation $T_\mathcal{D}(Y)$ when we use  a specified decoder $\mathcal{D}$ (as in Section 5). We have illustrated the channel model in Figure \ref{LevenshteinFig}. 

When we have an $e$-error-correcting code $C$, only substitution errors occur, and at most $t=e+\ell$ errors occur in a channel, then 
 \begin{equation}\label{TY}
 T(Y)=\bigcap_{\by\in Y} B_t(\by)\cap C.
 \end{equation}
  In this setup, the parameter $\LL$, i.e., the maximum size of $T(Y)$, has been studied in \cite{Levenshtein, yaakobi2018uncertainty, junnila2022levenshtein, junnila2020levenshtein}. For the case  $\LL=1$ see Theorem~\ref{LevL=1} below. However, when combinations of different error types are studied, much less is known \cite{Maria_Abu-Sini, Uusi_Maria_Abu-Sini, abu2023intersection} and even when only deletion or insertion errors occur we have many open problems depending on the code $C$ and the list size $\LL$ \cite{gabrys2018sequence, Maria_Abu-Sini, abu2023intersection, Uusi_Maria_Abu-Sini, goyal2022sequence, levenshtein2001efficient}. Decoding algorithms for substitution errors have been studied in \cite{Levenshtein, yaakobi2018uncertainty, Uusi_Maria_Abu-Sini} and for deletion and insertion errors in \cite{gabrys2018sequence, levenshtein2001efficient, Uusi_Maria_Abu-Sini}.

Similar problem has also been considered (sometimes under the name \textit{trace reconstruction}), when each error has an independent probability to occur, that is, the maximum number of errors in a channel is not fixed, in for example \cite{batu2004reconstructing, cheraghchi2020coded, viswanathan2008improved, chen2022near}. 

\medskip
 
 Next, we introduce the Levenshtein's result which gives the exact required number of channels $N$  to have $\LL=1$ for a $q$-ary Hamming space.
 As is usual, when $j>i$ or $j<0$, we have $\binom{i}{j}=0$.
 
 \begin{theorem}[\cite{Levenshtein}]\label{LevL=1}
 	Let $N\geq N_{t,e}$ and $C\subseteq \F_q^n$ be an $e$-error-correcting code. Then $\LL=1$ if
 	
 	$$N_{t,e}=\sum_{i=0}^{\ell-1}\binom{n-2e-1}{i}(q-1)^i\sum_{k=e+i-(\ell-1)}^{t-i}\sum_{j=e+i-(\ell-1)}^{t-i}\binom{2e+1}{k}\binom{2e+1-k}{j}(q-2)^{2e+1-k-j}+1.$$
 	
 	For $e=0$ there exists a simplified version of the bound above
 	\begin{equation}\label{simpleLev}
 	N_{t,0}= q\sum_{i=0}^{t-1}\binom{n-1}{i}(q-1)^i+1.
 	\end{equation}
 	\end{theorem}
 In particular, notice that if $N\geq N_{t,e}$, then $T(Y)=\{\bx\}$ for any transmitted $\bx$ and corresponding sets of output words $Y$. Further note, that when $C=\F^n_q$, we have $e=0$.

In this article, we concentrate on generalizing the results on the Levenshtein's reconstruction problem in the binary case to a more general $q$-ary environment. Especially the case with $q=4$ is relevant for DNA-memories, since the information is encoded in them using the four different types of nucleotides~\cite{Uusi_Maria_Abu-Sini}, although even larger alphabet sizes are of interest for polymer-based memory systems~\cite{laurent2020high,MR3376124}.

The structure of this paper is as follows.  In Section~\ref{SecErasure}, we consider channels with erasure errors and determine the list size $\LL$ (including $\LL=1$) with respect to the number $N$ of channels. Notice that the introduction of erasure errors to a word of $\Z_q^n$ can be interpreted as changing the underlying $q$-ary alphabet to one with $q+1$ symbols (including the erasure symbol $*$). In the next section, we allow a channel to have two types of errors simultaneously, namely, substitution errors and erasure errors. We provide the exact number of channels required in $q$-ary case to  reconstruct the transmitted word uniquely (that is, $\LL=1$). In Section \ref{SecqlistComb}, we consider only substitution errors and determine the \emph{exact} number of channels to guarantee that the list size $\LL$ of possible transmitted words is always a \emph{constant} with respect to the length $n$. This result is a $q$-ary generalization of a binary result introduced in~\cite{junnila2020levenshtein}.  We note that the results in this section also have independent theoretical interest as we give the exact size of the intersection of some $q$-ary Hamming balls. Previously, in~\cite{Uusi_Maria_Abu-Sini}, an efficient majority algorithm has been introduced for decoding the transmitted codeword in the binary case. 
We generalize the previous algorithm for larger $q$ in Section ~\ref{Sec:DecodingQ}. We also consider a new \emph{list}-decoding algorithm for determining $T(Y)$. Here we restrict ourselves to the binary alphabet since the algorithm is based on some rather complicated results, which are known only in the binary case. In Section~\ref{SecLargeq}, we consider the likelihood of $|T(Y)|=1$ for large $q$ when the error type in a channel is, in turn, insertion, substitution, deletion or erasure error. In that section, we consider $Y$ as a multiset while everywhere else in the paper $Y$ is a set. We observe that for large $q$ the insertion, substitution and deletion errors behave in different ways.  In particular, the result for insertion error is rather surprising. Finally, we conclude in Section~\ref{SecConclusion}.

\section{Erasures occurring in the channels}\label{SecErasure}
In this section, we consider erasure errors.  This  may also be understood as replacing a $q$-ary Hamming space by a $(q+1)$-ary Hamming space and by considering substitute errors where some symbols are substituted by $q$. From this perspective, we consider the symbol $q$ (or $\eras$) to belong to the support. Similarly, we consider each symbol $q$ to increase the weight of the word by one. 
Furthermore, we define the Hamming distance of two words $\bw,\bz$ which may contain erasures from this perspective, that is, when calculating the distance, if $w_i=q$ and $z_i\neq q$, then this difference increases the distance of $\bw$ and $\bz$ by one. Furthermore, observe that the size of $t$-radius erasure ball in $\F_q^n$ is $V_2(n,t)$. In other words,  when  at most $t$ erasures may occur in a channel, the number of channels is at most $V_2(n,t)$.

Let us denote by $A_q(n,d)$ the cardinality of a maximum size code $C\subseteq \F_q^n$ with minimum distance $d$. Note that when $d>n$, we trivially have $A_q(n,d)=1$. 
In general, there is no closed formula for $A_q(n,d)$, but numerous bounds and exact values are known for it (see \cite{HandBookCT}).

\begin{theorem}\label{ThmErasureExact} Assume that $n\geq t$ and at most $t$ erasures may occur in a channel.
	\begin{itemize}
		\item[(i)]	Let $C=\F^n_q$.  We have the maximum list size $\LL=q^a$ if the number of channels $N$ satisfies $$V_2(n-a-1,t-a-1)+1\leq N\leq V_2(n-a,t-a)$$ for some integer   $a\in [0,t-1]$. In particular, we get $\LL=1$ when $N\geq V_2(n-1,t-1)+1$.
		\item[(ii)] 	Let $C\subseteq\F^n_q$ have minimum distance $d$. We have $\LL=A_q(a,d)$ if $$V_2(n-a-1,t-a-1)+1\leq N\leq V_2(n-a,t-a)$$ for some integer $a\in [1,t-1]$.
	\end{itemize}

\end{theorem}

\begin{proof}  Observe that if we have output words $\by_i\in Y$ $(i=1,\dots,N)$ such that at least one of them does not have an erasure in the $i$th coordinate, then we can deduce what that symbol was in the transmitted word $\bx$.
	
	(i) Let us have $V_2(n-a-1,t-a-1)+1\leq N\leq V_2(n-a,t-a)$ for some integer $a\in [0,t-1]$.
	
	Consider an output word set such that each $\by\in Y$ has erasures in each coordinate of $[1,a]$. There are $V_2(n-a,t-a)$ possibilities for such words when at most $t$ erasures occur. Clearly there are $q^a$ possibilities for possible transmitted words in this case as $C=\F^n_q$. Thus, $\LL\geq q^a$. Let us then show that $\LL\leq q^a$.
	
	Suppose to the contrary that $\LL>q^a$. This is possible only if there exists such an output word set $Y$ that, for some set $S$ of $a+1$ coordinates, each $\by\in Y$ has the symbol $\eras$ in all of these coordinates. Now each output word has $a+1$ erasures in the coordinate set $S$ and at most $t-a-1$ erasures in coordinates $[1,n]\setminus S$. Thus, $|Y|\leq V_2(n-a-1,t-a-1)<N$. Hence, the claim follows.
	
	(ii) This case goes analogously. 	Let us have $V_2(n-a-1,t-a-1)+1\leq N\leq V_2(n-a,t-a)$ for some
		integer $a\in [1,t-1]$. 	Consider an output word set such that each $\by\in Y$ has erasures in the symbols within coordinate positions $[1,a]$. 
		 There are $V_2(n-a,t-a)$ possibilities for such words when at most $t\geq a$ erasures occur. Moreover, there are at most $A_q(a,d)$ possible transmitted codewords as the underlying code has minimum distance $d$. Thus, $\LL\geq A_q(a,d)$. Let us then show that $\LL\leq A_q(a,d)$.
		
		Suppose to the contrary that $\LL>A_q(a,d)$. This is possible only if there exists such an output word set $Y$ that, for some set $S$ of at least $a+1$ coordinates, each $\by\in Y$ has the symbol $\eras$ in all of these coordinates. Now each output word has $a+1$ erasures in coordinate set $S$ and at most $t-a-1$ erasures in coordinates $[1,n]\setminus S$. Thus, $|Y|\leq V_2(n-a-1,t-a-1)<N$. Hence, the claim follows.
	\end{proof}
	
	In particular, when we compare Case (i) of the previous theorem for erasures with $\LL=1$ to Theorem \ref{LevL=1} with code $C=\F^n_q$ (and hence $e=0$), we observe that unlike in the case of \emph{substitution}, the number of required channels does not increase as $q$ increases. Furthermore, when we consider a binary case $q=2$, we notice (see \eqref{simpleLev}) that substitution errors require approximately twice as many channels as erasure errors, that is, $2V_2(n-1,t-1)+1$ channels.


%
%

\section{Substitutions and erasures occurring in the channels} \label{SecErasureSubstitution}

In this section, we consider the case where at most $t_s$ substitutions and $t_e$ erasures occur in any channel. Note that we do not define the order in which these different types of errors may occur as it will not affect in our results. Moreover, when the underlying code $C$ is $e$-error-correcting, we denote $t_s=e+\ell$. We denote $$V_q(n,a_1,a_2)=\sum_{i=0}^{a_1}\binom{n}{i}\sum_{j=0}^{a_2}\binom{n-i}{j}(q-1)^j.$$ Observe that value $V_q(n,a_1,a_2)$ denotes in how many different ways we may
 assign at most $a_1$ erasures and $a_2$ substitutions to a word in a $q$-ary Hamming space, that is, $|B^{e,s}_{t_e,t_s}(\bx)|=V_q(n,t_e,t_s)$.
 We denote by  $d_e(\bx,\by)=d(\bx,\by)-|\{i\mid x_i=\eras \text{ xor } y_i=\eras\}|$, that is,  $d_e$ denotes the number of coordinates with symbols other than $\eras$ in which $\bx$ and $\by$ differ.

First, we recall a result from Levenshtein for intersections of substitution balls. Note that Levenshtein required that $d(\bx,\bx')\leq 2t$. However, when this is not true, the intersections of balls are simply empty and the statement still holds.

\begin{lemma}[Corollary 1 of \cite{Levenshtein}]\label{LemSubstMonotone}
Let $\bx_1,\bx_2,\bx_3,\bx_4\in \F_q^n$ and $d(\bx_1,\bx_2)\leq d(\bx_3,\bx_4)$. We have $|B_t(\bx_1)\cap B_t(\bx_2)| \geq |B_t(\bx_3)\cap B_t(\bx_4)|$.
\end{lemma}

In the following lemma, we generalize Levenshtein's result for combinations of substitution and erasure errors.

\begin{lemma}\label{LemErasureBalls}
Let $\bx_1,\bx_2,\bx_3,\bx_4\in \F^n_q$ be such that $d(\bx_1,\bx_2)\leq d(\bx_3,\bx_4)$. We have $$|B^{e,s}_{t_e,t_s}(\bx_1)\cap B^{e,s}_{t_e,t_s}(\bx_2)|\geq |B^{e,s}_{t_e,t_s}(\bx_3)\cap B^{e,s}_{t_e,t_s}(\bx_4)|.$$
\end{lemma}
\begin{proof}
Notice that we are only interested in the cardinalities of intersections of balls. Moreover, the balls $B^{e,s}_{t_e,t_s}(\bx_1)$ or $B^{e,s}_{t_e,t_s}(\bx_2)$ do not interact with balls $B^{e,s}_{t_e,t_s}(\bx_3)$ or $B^{e,s}_{t_e,t_s}(\bx_4)$. Hence, we may assume without loss of generality that $\bx_1=\nolla=\bx_3$.  Observe that now value $|B^{e,s}_{t_e,t_s}(\bx_1)\cap B^{e,s}_{t_e,t_s}(\bx_2)|$ depends only on $w(\bx_2)$ due to the symmetries of Hamming space and the same holds for $\bx_3$ and $\bx_4$.  Hence, if $d(\bx_1,\bx_2)=d=d(\bx_3,\bx_4)$, then the claim follows. We prove the lemma for $d(\bx_1,\bx_2)=d$ and $d(\bx_3,\bx_4)= d+1$. The claim follows by applying this iteratively. 
Hence, we assume without loss of generality that $\supp(\bx_2)=\{1,\dots, d\}$ and $\supp(\bx_4)=\{1,\dots, d+1\}$.

Observe that if $\by \in B_{t_e, t_s}^{e,s}(\bx_1) \cap B_{t_e, t_s}^{e,s}(\bx_2)$, then the erasures must have occurred in the same coordinates in $\bx_1$ and $\bx_2$. 
Let us denote by $D(\be,\bw)$, where $\be\in \F_2^n$ and $\bw\in \F_q^n$, a word $\bw'=D(\be,\bw)\in \F_q^{n-w(\be)}$ obtained from $\bw$ by deleting the symbols on the coordinates defined by $\supp(\be)$. We note that 
$$|B^{e,s}_{t_e,t_s}(\bx_1)\cap B^{e,s}_{t_e,t_s}(\bx_2)|= \sum_{w(\be)\leq t_e,\be\in \F_2^n}|B_{t_s}(D(\be,\bx_1))\cap B_{t_s}(D(\be,\bx_2))|$$ 
and 
$$|B^{e,s}_{t_e,t_s}(\bx_3)\cap B^{e,s}_{t_e,t_s}(\bx_4)|= \sum_{w(\be)\leq t_e,\be\in \F_2^n}|B_{t_s}(D(\be,\bx_3))\cap B_{t_s}(D(\be,\bx_4))|.$$

Notice that we have $d(D(\be,\bx_1),D(\be,\bx_2))\leq d(D(\be,\bx_3),D(\be,\bx_4))$ for each $\be\in  \F^n_2$. Hence, by Lemma \ref{LemSubstMonotone}, we have $$|B_{t_s}(D(\be,\bx_1))\cap B_{t_s}(D(\be,\bx_2))|\geq |B_{t_s}(D(\be,\bx_3))\cap B_{t_s}(D(\be,\bx_4))|.$$ Therefore, 
$|B^{e,s}_{t_e,t_s}(\bx_1)\cap B^{e,s}_{t_e,t_s}(\bx_2)| \geq |B^{e,s}_{t_e,t_s}(\bx_3)\cap B^{e,s}_{t_e,t_s}(\bx_4)|$, as claimed.
\end{proof}


\begin{theorem}\label{ThmEraSubExactMinDist}
	Let $n\geq t_e+t_s$ and $C$ be a code with minimum distance $d$. 
	 We have $\LL=1$ if \begin{align*}
		N\geq N'=&\sum_{i_e=0}^{t_e}\sum_{j_e=0}^{t_e-i_e}\binom{n-d}{i_e}\binom{d}{j_e}\sum_{i_1=0}^{t_s-\lceil(d-j_e)/2\rceil}\binom{n-d-i_e}{i_1}(q-1)^{i_1}\\
		\cdot&\sum_{i_2=d+i_1-j_e-t_s}^{t_s-i_1}\binom{d-j_e}{i_2}
		\sum_{i_3=0}^{t_s-i_1-i_2}\binom{d-j_e-i_2}{i_3}(q-2)^{i_3}+1\\
		=&V_q(n,t_e,t_s-d)+\sum_{i_e=0}^{t_e}\sum_{j_e=0}^{t_e-i_e}\binom{n-d}{i_e}\binom{d}{j_e}\sum_{i_1=0}^{t_s-\lceil(d-j_e)/2\rceil}\binom{n-d-i_e}{i_1}(q-1)^{i_1}\\
		\cdot&\sum_{i_2=d+i_1-j_e-t_s}^{t_s-i_1}\binom{d-j_e}{i_2}
		\sum_{i_3=t_s+1-i_1-i_2-d}^{t_s-i_1-i_2}\binom{d-j_e-i_2}{i_3}(q-2)^{i_3}+1.\\
	\end{align*}
	If $N<N'$, then $\LL\geq2$.
\end{theorem}
\begin{proof}
	%
	%
	Let $d(\bx,\bx')=d'\geq d$ for some $\bx,\bx'\in \F^n_q$. We may assume, without loss of generality, that $\bx=\nolla$. Thus, $w(\bx')=d'$. Consider $B=B^{e,s}_{t_e,t_s}(\bx)\cap B^{e,s}_{t_e,t_s}(\bx')$. To maximize the size of  $B$, we assume by Lemma \ref{LemErasureBalls} that $d'=d$. 
	The number of channels $N'$ which is required for $\LL=1$ equals $N'=|B|+1$. Furthermore, if $N\leq |B|$, then we have $\LL\geq2$ as we can have $\bx$ and $\bx'$ in $T(Y')$ for $Y'\subseteq B$. Hence, it is enough to show that $N'=|B|+1$.	

\begin{figure}[htp]	 \centering%
\begin{tikzpicture}
\node[align=left] at (-2.05,1.05) {$i_2$};
\node[align=left] at (-0.74,1.05) {$i_3$};
\node[align=left] at (0.0,1.05) {$j_e$};
\node[align=left] at (0.75,1.05) {$i_1$};
\node[align=left] at (1.51,1.05) {$i_e$};
\node[align=left] at (-1.07,-0.9) {$d$};
\node[align=left] at (0,0) {
$\bx$  \hspace{-0.13em} $\overbrace{0 \hspace{0.334em}0}$  0 0  $\overbrace{0 \hspace{0.334em}0}$ $\overbrace{0\hspace{0.334em} 0}$  $\overbrace{0\hspace{0.334em} 0}$  $\overbrace{0\hspace{0.334em} 0}$  0 ... 0\\
$\bx'$ \hspace{-0.18em} 1 1\hspace{0.25em} 1 1\hspace{0.25em} 1 1\hspace{0.46em} 1 1\hspace{0.46em} 0 0\hspace{0.46em} 0 0\hspace{0.25em} 0 ... 0\\
 
$\y$ \hspace{0.1em} $\underbrace{1 \hspace{0.334em} 1 \hspace{0.6em} 0 \hspace{0.334em} 0\hspace{0.6em} 2 \hspace{0.334em} 2\hspace{0.52em} \eras \hspace{0.13em} \eras}$\hspace{0.45em} 1 1\hspace{0.46em} $\eras$ $\eras$ \hspace{-0.08em} 0 ... 0\\
};
\end{tikzpicture}	
\caption{Transmitted word candidates $\bx$ and $\bx'$ at distance $d$ from each other together with output word $\by$ and variables $i_1,i_2,i_3,i_e$ and $j_e$ as in the proof of Theorem \ref{ThmEraSubExactMinDist}. Note that due to the symmetries of Hamming space, we may assume without loss of generality that $\bx=\nolla$ and $\bx'$ contains only $0$'s and $1$'s.
}\label{FigEraSubProof}
\end{figure}

	Consider next a word $\by\in B$. In the following discussions, we consider variables $i_1,i_2,i_3,i_e$ and $j_e$ together with words $\bx,\bx'$ and $\by$. For an example illustration of these variables, see Figure \ref{FigEraSubProof}. Let us denote by $j_e$ the number of erasures in $\by$ such that they occur within $\supp(\bx')$, let $i_1=|\supp(\by)\setminus(\supp(\bx')\cup\{i\mid y_i=\eras\})|$ and $i_2=|\vsupp(\by)\cap \vsupp(\bx')|$. We have
	  \begin{equation}\label{de}
	  	d_e(\by,\bx')=(d-i_2-j_e)+i_1\leq t_s.
	  \end{equation}	
	  Indeed, $w(\bx')=d$ and each element in $\vsupp(\bx')$ which is not contained in $\vsupp(\by)$ or in which $\by$ does not have an erasure increases $d_e(\by,\bx')$ by one. This implies the term $d-i_2-j_e$. Furthermore, each element in $\supp(\by)\setminus\supp(\bx')$ which is not an erasure, increases $d_e(\by,\bx')$ by one. 
	  	 
	  	 As the total number of substitutions in $\by$ is at most $t_s$, we have $i_2\leq t_s-i_1$, since $d_e(\bx,\by)\leq t_s$, and $i_1$ is an integer. 	 
	  	By Equation (\ref{de}), we have $i_1\leq t_s+i_2-(d-j_e)\leq 2t_s-i_1-(d-j_e)$. Thus,  \begin{equation}\label{eqi1raja}
	  	i_1\leq t_s-\lceil(d-j_e)/2\rceil.
	  	\end{equation} Moreover,  using \eqref{de} we get \begin{equation}\label{eqi2raja}
	  	i_2\geq d-j_e-t_s+i_1.\end{equation} 
	 	  Finally, let us denote by $i_e$ the number of erasures, in $\by$, outside of coordinates of $\supp(\bx')$ and by $i_3$ the number of substitution errors, in $\by$, in $\supp(\bx')$ such that the values of $\by$ and $\bx'$ differ within those coordinates (hence there are $(q-2)^{i_3}$ ways to choose them). In other words, $i_3=|\supp(\bx')\cap\supp(\by)\cap\{i\mid y_i\neq \eras \text{ and } x_i'\neq y_i\}|$. Note that since the total number of substitutions in $\by$ is at most $t_s$, we have \begin{equation}\label{eqi3raja}
	 	  0\leq i_3\leq t_s-i_1-i_2.
	 	  \end{equation} 
	 	  
	 	  Assume next that word $\by$ is such that above Inequalities (\ref{eqi1raja}), (\ref{eqi2raja}) and (\ref{eqi3raja}) hold for it. Together with lower bound $0\leq i_1$, upper bound $i_2\leq t_s-i_1$, $0\leq i_e\leq t_e$ and $0\leq j_e\leq t_e-i_e$ we show that then $\by\in B$. First of all,  $\by$ is obtained with $i_1+i_2+i_3$ substitutions and $i_e+j_e$ erasures from $\bx$. Since  $i_1+i_2+i_3\leq t_s$ and $i_e+j_e\leq t_e$, we have $\by\in B^{e,s}_{t_e,t_s}(\bx)$. Furthermore, we may obtain $\by$ from $\bx'$ by first placing the $i_e+j_e\leq t_e$ erasure symbols $\eras$ to corresponding coordinates, then executing $i_1$ suitable substitutions to coordinates outside of $\supp(\bx')$ and finally by making $i_3$ suitable substitutions to the coordinates in $\supp(\bx')$. Since $i_1+i_3\leq t_s$, we have $\by\in B^{e,s}_{t_e,t_s}(\bx')$ and thus, $\by\in B$ as claimed.
	 	  
	 	  Therefore, we have

	\begin{align*}
		|B|=&\sum_{i_e=0}^{t_e}\sum_{j_e=0}^{t_e-i_e}\binom{n-d}{i_e}\binom{d}{j_e}\sum_{i_1=0}^{t_s-\lceil(d-j_e)/2\rceil}\binom{n-d-i_e}{i_1}(q-1)^{i_1}\\
		\cdot&\sum_{i_2=d+i_1-j_e-t_s}^{t_s-i_1}\binom{d-j_e}{i_2}
		\sum_{i_3=0}^{t_s-i_1-i_2}\binom{d-j_e-i_2}{i_3}(q-2)^{i_3}\\
		=&V_q(n,t_e,t_s-d)+\sum_{i_e=0}^{t_e}\sum_{j_e=0}^{t_e-i_e}\binom{n-d}{i_e}\binom{d}{j_e}\sum_{i_1=0}^{t_s-\lceil(d-j_e)/2\rceil}\binom{n-d-i_e}{i_1}(q-1)^{i_1}\\
		\cdot&\sum_{i_2=d+i_1-j_e-t_s}^{t_s-i_1}\binom{d-j_e}{i_2}
		\sum_{i_3=t_s+1-i_1-i_2-d}^{t_s-i_1-i_2}\binom{d-j_e-i_2}{i_3}(q-2)^{i_3}.\\
	\end{align*}
	
	The latter equality above follows from the observations 1) that $\by\in B$, if  $d_e(\nolla,\by)\leq t_s-d$ and at most $t_e$ erasures occur to $\by$. Indeed, we may transform $\bx'$ into $\nolla$ with $d$ substitutions and then $\nolla$ into $\by$ with $t_s-d$ substitutions and at most $t_e$ erasures. Hence, $B_{t_e,t_s-d}^{e,s}(\nolla)\subseteq B$. Furthermore, we observe 2) that the last sum term considers exactly the cases where at least $t_s-d+1$ substitutions occur since in the last sum term we have $i_3\geq t_s+1-i_1-i_2+d$ and hence, $i_1+i_2+i_3\geq t_s-d+1$. Hence, the sum does not consider the output words already considered in $B_{t_e,t_s-d}^{e,s}(\nolla)\subseteq B$.
	Since $N'=|B|+1$, we have $\LL=1$.
\end{proof}

\section{A tight bound for constant \texorpdfstring{$q$}{q}-ary list size}\label{SecqlistComb}

In this section, we consider only \emph{substitution} errors and our aim is to find out the exact number of channels which guarantees that the list size is always a \emph{constant} with respect to the length $n$. We note that our results are also independently interesting from theoretical perspective as they give new information about the size of the intersection of $q$-ary Hamming balls. This perspective is briefly discussed at the end of the section.
We first show that if we have too few channels, then  there are $e$-error correcting codes in $\F_q^n$ such that the list size grows when $n$ increases.
Indeed, assume that the number of channels satisfies $N\leq V_q(n,\ell-1)$ and that all the output words in $Y\subseteq \F_q^n$ are located inside the ball $B_{\ell-1}(\nolla)$. Clearly, any word of weight $e+1$ in $\F_q^n$ belongs
to the intersection $\bigcap_{\by\in Y} B_t(\by)$. If our $e$-error-correcting code is such that each codeword has weight $e+1$, then all the codewords are in the intersection.
Such codes (usually called \emph{$q$-ary constant-weight codes}) have been widely studied in the literature (see, for example, \cite{Yeow} and references therein). For our purposes (since we are only interested in the dependence on $n$ and not on the largest possible constant-weight codes in $\F_q^n$), it is enough to choose  the code $C_e^n$ which consists of the words with the following value supports
$$\vsupp(\bc_j)=\{(j(e+1)+k,1)\mid k=1,2,\dots,e+1\}$$
where $j=0,1,\dots, \lfloor n/(e+1)\rfloor.$ (For example, if $n=7$ and $e=2$, then the code $C_2^7=\{1110000,0001110\})$. Since the minimum distance of $C_e^n$ equals $2e+2$ by \eqref{vsupp}, we obtain the following result.

\begin{theorem}\label{nonconstant}
Let $t=e+\ell$. If $N\leq V_q(n,\ell-1)$ then there exists an $e$-error-correcting code  in $\F_q^n$ such that  $$\LL\ge \lfloor n/(e+1)\rfloor. $$
\end{theorem}

In what follows, we show that if the number of channels is  larger than $V_q(n,\ell-1)$ (considered in Theorem~\ref{nonconstant}), that is, $N\geq V_q(n,\ell-1)+1$, then, given \emph{any} $e$-error-correcting code, the list size  is immediately independent of $n$ (when $n$ is large enough). Moreover, we will show that the constant upper bound, which we obtain in Theorem~\ref{Listapituus} for the list size, is actually \emph{optimal} as will be seen in Theorem~\ref{ell(q-1) list}. 

It should be noticed that although the Lemmas~\ref{l-1 kaukana} and \ref{Listasanat lahella} are rather similar to the binary case discussed in \cite{ junnila2020levenshtein}, our new techniques in Lemma~\ref{Keskussana} and Theorem~\ref{Listapituus} differ significantly from the ones in \cite{ junnila2020levenshtein} as we consider the more complicated case of  $q$-ary alphabet for $q>2$.

In the following lemma, we show that if $n$ and $N$ are large enough, then for a set of coordinate positions $\overline{D}$ of size $b$ and any codeword $\bc$, there exists an output word $\by$ such that $\by$ differs from $\bc$ in at least $\ell-1$ coordinate positions outside of $\overline{D}$.

\begin{lemma}\label{l-1 kaukana}
Assume that $Y\subseteq \F_q^n$, $|Y| = N \geq V_q(n,\ell-1)+1$, $C$ is an $e$-error-correcting code and $b$ is a positive integer. If 
$n\geq  \ell-2+(\ell-1)^22^b(q-1)^{b-1}$, 
then for any codeword $\bc \in T(Y)$ and for any set $\overline{D}\subseteq [1,n]$ with $|\overline{D}|=b$, there exists a word $\y\in Y$ such that $$|\supp(\bc-\y)\setminus \overline{D}|\geq \ell-1.$$
\end{lemma}
\begin{proof}
Let $\overline{D}\subseteq [1,n]$ and $|\overline{D}|=b$ for some fixed $b$. Without loss of generality, we may assume that $\bc = \nolla$. 
Suppose to the contrary that there does not exist a word $\y \in Y$ such that $|\supp(\bc - \y)\setminus \overline{D}| = |\supp(\y)\setminus \overline{D}|\geq \ell-1$, i.e., $|\supp(\y)\setminus \overline{D}| < \ell-1$ for all $\y \in Y$. Note that since $b>0$, we have $n\geq 2(\ell-2)$. This implies that the number of words in $Y$ is at most
\begin{align*}
&\sum_{j=0}^{\ell-2}\, \sum_{i=0}^{\min\{b,t-j\}}(q-1)^{i+j}\binom{b}{i}\binom{n-b}{j}\\
\leq&\sum_{j=0}^{\ell-2}\sum_{i=0}^{b}(q-1)^{i+j}\binom{b}{i}\binom{n-b}{j}\\
\leq&2^b(q-1)^b\sum_{j=0}^{\ell-2}(q-1)^j\binom{n-b}{j}\\
\leq& (\ell-1)2^b(q-1)^{b+\ell-2}\binom{n}{\ell-2}\\
=&  (\ell-1)\binom{n}{\ell-1}\frac{\ell-1}{n-\ell+2}2^b(q-1)^{b+\ell-2}\\
\leq&(q-1)^{\ell-1}\binom{n}{\ell-1}\\
<&V_q(n,\ell-1)\text,
\end{align*}
when $n\geq \ell-2+(\ell-1)^22^b(q-1)^{b-1}$. This contradicts with the assumption that $N = |Y| \geq V_q(n,\ell-1)+1$. Thus, the claim follows.
\end{proof}

In the following lemma, we show that if $n$ is large enough and $N\geq V_q(n,\ell-1)+1$, then the pairwise distances of codewords in $T$ are rather small.
\begin{lemma}\label{Listasanat lahella}
Let $n\geq\ell-2+(\ell-1)^22^{2t}(q-1)^{2t-1}$, $C$ be an $e$-error-correcting code and $|Y|=N\geq V_q(n,\ell-1)+1$. Then we have $d(\bc_1,\bc_2)\leq 2e+2$ for any  $\bc_1,\bc_2\in T(Y)$.
\begin{proof}
Let $\bc_1$ and $\bc_2$ be codewords in $T(Y)$. Without loss of generality, we may assume that $\bc_1=\nolla$. In order to show that $d(\bc_1,\bc_2)\leq 2e+2$, we suppose to the contrary that $d(\bc_1,\bc_2)\geq 2e+3$, i.e., $w(\bc_2)\geq 2e+3$. Since $\bc_1,\bc_2\in T(Y)$, we have $w(\bc_2) = d(\bc_1,\bc_2)\leq 2t$. Hence, there exists a set $\overline{D}\subseteq [1,n]$ such that $|\overline{D}|=2t$ and $\supp(\bc_2)\subseteq \overline{D}$.

Since $n\geq\ell-2+(\ell-1)^22^{2t}(q-1)^{2t-1}$, by Lemma \ref{l-1 kaukana}, there exists an output $\y\in Y$ such that $|\supp(\y)\setminus \supp(\bc_2)|\geq \ell-1$. Since $w(\y) = d(\y,\bc_1)\leq t$, we have $|\supp(\bc_2)\cap\supp(\y)|\leq e+1$; indeed, if $|\supp(\bc_2)\cap\supp(\y)|\geq e+2$, then $w(\y) = |\supp(\bc_2)\cap\supp(\y)| + |\supp(\y)\setminus \supp(\bc_2)| \geq (e+2) + (\ell-1) = t+1$ (a contradiction). This further implies that 
\begin{align*}
d(\bc_2,\y) &\geq (w(\bc_2)-|\supp(\bc_2)\cap \supp(\y)|)+\ell-1\\
&\geq  (2e+3-(e+1))+\ell-1 = t+1 \text.
\end{align*} This leads to a contradiction, and the claim follows.\end{proof}\end{lemma}

\begin{lemma}\label{Keskussana}
Let $\ell\geq1,q\geq3$, $C$ be an $e$-error-correcting code and $|Y|=N\geq V_q(n,\ell-1)+1$. Moreover, let $\bw\in \F^n_q$ be such a word that  $d(\bw,\bc)\leq e+1$ for each $\bc\in T(Y)$. 
Then we have $$|T(Y)|\leq \ell(q-1)+1.$$ 
\end{lemma}
\begin{proof}
We may assume without loss of generality that $\bw=\nolla$. 
Since $d(\bw,\bc)\leq e+1$ for each $\bc\in T(Y)$, we have $w(\bc)\leq e+1$.
Thus, since $C$ is an $e$-error-correcting code with minimum distance $2e+1$, if $|T(Y)|>1$, then $w(\bc) \in \{e, e+1\}$ for each $\bc \in T(Y)$ and there is at most one codeword of weight $e$, denoted by $\bc_0$. Since $N\geq  V_q(n,\ell-1)+1$, there is at least one output word of weight at least $\ell$. Let us say that for some $\by\in Y$ we have $w(\by)=\ell-1+a$ where  $1\leq a$. Observe that $a\leq 2e+2$ since $d(\by,\bc)\leq t$ for each $\bc\in T(Y)$.

Let us consider $\bc\in T(Y)$ with $w(\bc)=e+1$. Since $d(\by,\bc)\leq t$ for each $\bc\in T(Y)$ and, by \eqref{vsupp}, $d(\by,\bc)=w(\by)+w(\bc)-|\supp(\by)\cap \supp(\bc)|-|\vsupp(\by)\cap \vsupp(\bc)|=t+a-|\supp(\by)\cap \supp(\bc)|-|\vsupp(\by)\cap \vsupp(\bc)|$, we have $a\leq |\supp(\by)\cap \supp(\bc)|+|\vsupp(\by)\cap \vsupp(\bc)|$. Hence, we have $|\supp(\by)\cap \supp(\bc)|\geq \lceil\frac{a}{2}\rceil$. Moreover, since $C$ is $e$-error-correcting, we have $|\vsupp(\bc)\cap \vsupp(\bc')|=0$ for each $\bc'\in T(Y)$ other than $\bc$.   Thus, we have  at most $1+\frac{w(\by)(q-1)}{\lceil\frac{a}{2}\rceil}$ codewords in $T(Y)$. Indeed, we have $$|T(Y)\setminus\{\bc_0\}| \cdot \lceil a/2 \rceil \leq \sum_{\bc \in T(Y)} |\supp(\by) \cap \supp(\bc)| \leq w(\by)(q-1).$$
Here the first inequality is due to $|\supp(\by)\cap \supp(\bc)|\geq \lceil\frac{a}{2}\rceil$ and the second one follows from $|\vsupp(\bc)\cap \vsupp(\bc')|=0$.



 Notice that when $a=1$, this gives the claimed value $\ell(q-1)+1$.  Next we consider separately the cases where $a$ is odd and even.

Let us first consider the case where $a\geq1$ is odd and $\ell\geq2$. In that case, we have $$1+\frac{w(\by)(q-1)}{\lceil\frac{a}{2}\rceil}=1+\frac{2(\ell-1+a)(q-1)}{a+1}= 1 + 2 \cdot \frac{(\ell-2)+(a+1)}{a+1} \cdot (q-1) \leq 1 + \ell(q-1).$$ 

Let us then consider the case with even $a\geq4$ and $\ell\geq3$. We have $$1+\frac{w(\by)(q-1)}{\lceil\frac{a}{2}\rceil}=1+\frac{2(\ell-1+a)(q-1)}{a}\leq  1+\frac{(\ell+3)(q-1)}{2}\leq 1+\ell(q-1).$$ 

Hence, we are left with the cases 1) $a=2$ and $\ell\geq1$,  2) $a\geq4$ even and $\ell=2$, and 3) $a\geq3$ and $\ell=1$. Let us next concentrate on Case 1) $a=2$ and $\ell\geq2$. Now, $w(\by)=\ell+1$ and for $\bc\in T(Y)$ with $w(\bc)=e+1$, we have $2\leq |\supp(\by)\cap \supp(\bc)|+|\vsupp(\by)\cap \vsupp(\bc)|$. Thus, $|\supp(\by)\cap \supp(\bc)|\geq2$ or  $|\vsupp(\by)\cap \vsupp(\bc)|\geq1$. Since codewords in $T(Y)$ do not have overlapping value supports, there are at most $\ell+1$ words in $T(Y)$ such that $|\vsupp(\by)\cap \vsupp(\bc)|\geq1$. If the single word $\bc_0\in T(Y)$ with weight $e$ exists, then it has $|\supp(\by)\cap \supp(\bc_0)|\geq1$ since $d(\bc_0,\by)\leq t$. Let us denote by $k$ the number of codewords $\bc\in T(Y)$ with $|\supp(\by)\cap \supp(\bc)|\geq2$ and by $k'$ the number of codewords  $\bc\in T(Y)$ with $|\supp(\by)\cap \supp(\bc)|=1$. Then, we have $\sum_{\bc\in T(Y)}|\supp(\by)\cap \supp(\bc)|\geq k'+2k$. Moreover, $k'\leq \ell+2$ as at most $\ell+1$ words may have intersecting value support with $\by$ and the word $\bc_0$ may have its support intersecting $\by$ without having an intersecting value support. Furthermore, since none of these words have overlapping value supports, we have $\sum_{\bc\in T(Y)}|\supp(\by)\cap \supp(\bc)|\leq (\ell+1)(q-1)$. Hence, we get $k'+2k\leq (\ell+1)(q-1)$.
 To maximize $k'+k$, we may assume that $k'$ is maximal, that is, $k'=\ell+2$. Thus,
\begin{align*}
(\ell+2) + 2k=k'+2k&\leq (\ell+1)(q-1) \quad\iff \quad k \leq (\ell+1)(q-1)/2-\ell/2-1. 
\end{align*}
Hence, $k'+k\leq (\ell+1)(q-1)/2+\ell/2+1=\ell(q-1)/2+(q-1)/2+\ell/2+1$. For $\ell=1$, this gives upper bound $\ell(q-1)+3/2$ which implies $k'+k\leq 1+\ell(q-1)$ as $k'+k$ is integer. Moreover, we notice that the value is at most $\ell(q-1)+1$ also for $\ell\geq2$. Indeed, $(q-1)/2+\ell/2=(q-1+\ell)/2\leq (q-1)\ell/2$ and hence, $k'+k\leq \ell(q-1)/2+(q-1)\ell/2+1\leq \ell(q-1)+1$. 

Let us next consider Case 2) $\ell=2$ and $a\geq4$ is even. We have $w(\by)=a+1$. Recall that for each $\bc\in T(Y)$ with weight $e+1$, we have $|\supp(\bc)\cap \supp(\by)|\geq \lceil a/2\rceil=a/2$ and this lower bound is attained only when  $|\vsupp(\bc)\cap \vsupp(\by)|=a/2$. Indeed, if $|\vsupp(\bc)\cap \vsupp(\by)|\leq a/2-1$, then $|\supp(\bc)\cap \supp(\by)|\geq a/2+1$. Let us denote by $k$ the number of codewords of weight $e+1$ with $|\supp(\bc)\cap \supp(\by)|\geq a/2+1$ and by $k'$ the  number of codewords of weight $e+1$ with $|\supp(\bc)\cap \supp(\by)|= a/2$. Note that $|\supp(\bc_0)\cap \supp(\by)|\geq a/2$. Recall that for any two codewords in $T(Y)$, we have $|\supp(\bc)\cap \supp(\bc')|\leq1$ and $|\vsupp(\bc)\cap \vsupp(\bc')|=0$. Hence, we have $k\leq 2$ and $k'\leq2$. Furthermore, the only codeword of weight $e$ which may exist is $\bc_0$ and we have $|T(Y)|\leq k+k'+1$. Hence, $k+k'+1\leq5 \leq 1+\ell(q-1)$ as $q\geq3$.

Let us finally consider Case 3) $\ell=1$ and $a\geq3$. We have $w(\by)=a$. Recall that for each $\bc\in T(Y)$ with weight $e+1$, we have $|\supp(\bc)\cap \supp(\by)|\geq \lceil a/2\rceil$ and this lower bound is attained only when $|\vsupp(\bc)\cap \vsupp(\by)|\geq \lfloor a/2\rfloor$. Similarly to the codewords of weight $e+1$ discussed in the second paragraph of the proof, we obtain that $|\supp(\bc_0)\cap \supp(\by)|\geq \lfloor a/2\rfloor$. Denote by $k$ the number of codewords with $|\supp(\bc)\cap \supp(\by)|\geq \lceil a/2\rceil+1$ and by $k'$ the number of codewords with  $|\supp(\bc)\cap \supp(\by)|\leq \lceil a/2\rceil$. When $a\geq5$ is odd, since we have  $|\supp(\bc)\cap \supp(\bc')|\leq1$, there can be at most two codewords with $|\supp(\bc)\cap \supp(\by)|\geq (a+1)/2$. Hence, we have $k+k'\leq3\leq 1+\ell(q-1)$, taking into account $\bc_0$. When $a=3$ and $q=3$, the codewords in $T(Y)$ have a total of six 'symbols' at three coordinate positions available for them since $(q-1)\cdot w(\by)=6$. As each codeword besides $\bc_0$ uses at least two of them (since $\lceil a/2\rceil=2$) and the value supports may not intersect, we have $k+k'\leq 3\leq \ell(q-1)+1$. When $a=3$ and $q\geq4$, we may have at most three codewords whose supports intersect with $\supp(\by)$ in addition to $\bc_0$ since $|\supp(\bc)\cap\supp(\bc')|\leq1$ for any two codewords. Hence, $k+k'\leq 4<\ell(q-1)+1$. 

When $a$ is even, supports of any two codewords of weight $e+1$ cover $\supp(\by)$. Indeed, since $|\supp(\bc)\cap \supp(\bc')|\leq1$ for any two codewords $\bc,\bc'$ of weight $e+1$, if codeword $\bc$ intersects $\by$ on $a/2+1$ coordinates, then together with another codeword intersecting $\by$ on at least $a/2$ coordinates, they cover $\supp(\by)$. If a codeword $\bc$ has $|\supp(\bc)\cap\supp(\by)|=a/2$, then $|\vsupp(\bc)\cap\vsupp(\by)|=a/2$. Since $|\vsupp(\bc)\cap\vsupp(\bc')|=0$ for two codewords $\bc,\bc'$ of weight $e+1$, they cover $\supp(\by)$. In conclusion, any two codewords of weight $e+1$ cover $\supp(\by)$.
 Hence, when $a\geq6$, there are at most two codewords in $T(Y)$ in addition to $\bc_0$ and $k+k'\leq 3$. Assume next $a=4$. We have $k\leq1$ since $|\supp(\bc)\cap \supp(\bc')|\leq1$ for any two codewords. Recall that if $|\supp(\bc)\cap \supp(\by)|=a/2$, then $|\vsupp(\bc)\cap \vsupp(\by)|=a/2$ for any codeword $\bc$ with $w(\bc)=e+1$.
If $k=1$ and $\bc_1$ is such that $|\supp(\bc_1)\cap \supp(\by)|\geq a/2+1= 3$, then we can have only one (other) codeword with $|\vsupp(\bc)\cap \vsupp(\by)|\geq 2$ since $|\vsupp(\bc')\cap \vsupp(\bc'')|=0$ and $|\supp(\bc')\cap \supp(\bc'')|\leq 1$ for each $\bc',\bc''\in T(Y)$. Hence, $k'\leq 2$ in this case and $k+k'\leq3$. Finally, if $k=0$, then $k'\leq 3$ since there can be at most two codewords whose value supports intersect $\vsupp(\by)$ at two coordinate positions. Hence, also in this case $k+k'\leq 3\leq 1+\ell(q-1)$.
%
%
%
%
%

Therefore, in each case we have $|T(Y)|\leq \ell(q-1)+1$ and the claim follows.
\end{proof}

\begin{theorem}\label{Listapituus}
Let $n\geq\ell-2+(\ell-1)^22^b(q-1)^{b-1}$, $C$ be an $e$-error-correcting code, $|Y|=N\geq V_q(n,\ell-1)+1$, $\ell\geq1$, $q\geq3$ and $b\geq (\ell(q-1)+1)(2e+2)$. Then we have $\LL\leq \ell(q-1)+1$.
\end{theorem}
\begin{proof}
Let us denote  $T(Y)=\{\bc_0,\bc_1,\dots,\bc_{|T(Y)|-1}\}$.
Moreover, let us assume without loss of generality that $\bc_0=\nolla$. We have $2e+1\leq d(\bc_i,\bc_j)\leq2e+2$ for each distinct pair of indices $i,j\in[0,|T(Y)|-1]$ by Lemma \ref{Listasanat lahella} and the definition of $e$-correcting codes. Thus, $w(\bc_i)\leq 2e+2$ for each $i$. Suppose to the contrary that $|T(Y)|\geq\ell(q-1)+2$. Moreover, let us denote by $\overline{D}=\bigcup_{i=0}^{\ell(q-1)+1}\supp(\bc_i)$ 
and let $\by\in Y$ be an output word such that $|\supp(\by)\setminus \overline{D}|\geq\ell-1$. Word $\by$ exists by Lemma \ref{l-1 kaukana} since $n\geq\ell-2+(\ell-1)^22^b(q-1)^{b-1}$. 
Let us denote by $\s\in \F^n_q$ a word such that $\supp(\s)=\overline{D}\cap\supp(\by)$ and $\vsupp(\s)\subseteq \vsupp(\by)$. We have \begin{align*}
t\geq &d(\by,\bc_i)\geq\ell-1+d(\s,\bc_i)\iff
e+1 \geq d(\s,\bc_i).
\end{align*}

By Lemma \ref{Keskussana} together with choice $\bw=\s$, there are at most $\ell(q-1)+1$ codewords in $T(Y)$. This contradicts our assumption that $|T(Y)|\geq \ell(q-1)+2$. 
 Thus, $|T(Y)|\leq \ell(q-1)+1$. Since this holds for any $Y$ with $|Y|=N$, we have $\LL\leq \ell(q-1)+1$. 
\end{proof}

%
%

The next result shows that our bound on the list size is actually \emph{optimal} for $e$-error-correcting code when $N= V_q(n,\ell-1)+1$.

\begin{theorem}\label{ell(q-1) list}
	Let $n\geq e(q-1)\ell+t$ and $N\leq V_q(n,\ell-1)+1$. There exists such an $e$-error-correcting code that $$\LL\geq (q-1)\ell+1.$$
\end{theorem}
\begin{proof}
	Let us denote
		$C=\{\bc_1,\dots,\bc_{(q-1)\ell+1}\}$. For integers $h \in [0, \ell-1]$ and $p \in [1, q-1]$ let $i = h(q-1)+p$. For each $i \in [1, (q-1)\ell]$, we have

	$$\vsupp(\bc_i)=\{(h+1,p)\}\cup\{(j,1)\mid j\in[(i-1)e+\ell+1,i\cdot e+\ell]\}$$ 
	and $$\vsupp(\bc_{(q-1)\ell+1})=\{(j,1)\mid j\in[e(q-1)\ell+\ell+1,e(q-1)\ell+t]\}.$$ For example, if $e=\ell=2$, $q=3$ and $n=12=e(q-1)\ell+t$, then the code consists of the words \begin{align*}
	\bc_1=~&10~11~00~00~00~00,\\
	\bc_2=~&20~00~11~00~00~00,\\
	\bc_3=~&01~00~00~11~00~00,\\
	\bc_4=~&02~00~00~00~11~00,\\
	\bc_5=~&00~00~00~00~00~11.
	\end{align*}
	
	Let us now assume that $Y=\{\by\mid w(\by)\leq \ell-1\}\cup\{\by_\ell\}$ where $\vsupp(\by_\ell)=\{(j,1)\mid j\in[1,\ell]\}$. Observe that $|Y|=V_q(n,\ell-1)+1$.
	
	We claim that $C$ is an $e$-error-correcting code and $C\subseteq \bigcap_{\by\in Y}B_t(\by)$. Let us first consider the minimum distance of $C$. Observe, $w(\bc_{(q-1)\ell+1})=e$ while all other codewords have weight $e+1$. Furthermore, no codewords have intersecting value supports, supports of $(e+1)$-weight words intersect in at most one coordinate (within the first $\ell$ coordinates) and the support of the $e$-weight codeword does not intersect with the support of any other codeword. Thus, $C$ has minimum distance of $2e+1$ and is an $e$-error-correcting code.
	
	Let us then consider the distance between the codewords in $C$ and the output words in $Y$. Let $\by\in Y$. We have $d(\bc_i,\by)\leq w(\bc_i)+w(\by)$. Thus, when $w(\by)\leq \ell-1$, the distance is at most $t$. Hence, we only have to consider the distance between $\by_\ell$ and codewords of weight $e+1$. We immediately notice that $\by_\ell$ and those codewords have intersecting supports. So the distance is at most $d(\bc_i,\by_\ell)\leq w(\bc_i)+w(\by_\ell)-1=t+1-1=t$. Hence, the claim follows.
\end{proof}

We note that give large enough $n$, $\ell\geq1$, $q\geq3$ and $\ell(q-1)+1$ words $\bc\in W$ with minimum distance $2e+1$, then the intersection of $t$-radius Hamming balls centred at words in $W$ has cardinality of at most $$\bigcap_{\bc\in W}|B_t(\bc)|\leq V_q(n,\ell-1)+1.$$

\section{Decoding algorithm for substitution errors}\label{Sec:DecodingQ}

We describe the (well-known) majority algorithm in the $q$-ary Hamming space for substitution errors using similar terminology and notation as in~\cite{Uusi_Maria_Abu-Sini}. In this section, we assume that $q\geq3$ and $e\geq1$. Previously a decoding algorithm of optimal complexity has been presented for all values of $q$ and $e=0$ in \cite{levenshtein2001efficient}, for $q=2$ and $e=1$ in \cite{yaakobi2018uncertainty} and for $q=2$ and all values of $e$ in \cite{Uusi_Maria_Abu-Sini}.

We denote the coordinates of the output words $\by_j \in Y$ by $\by_j = (y_{j,1}, y_{j,2}, \ldots, y_{j,n})$. Furthermore, the number of symbols $j$ in the $i$th coordinates of the output words are respectively denoted by
\[
m_{i,j} = |\{k \in \{1, 2, \ldots, N\} \mid y_{k,i} = j\}|.
\]
Moreover, in the $i$th coordinate, the number of occurrences of the most common value $j$ is denoted by $M(i)=m_{i,j}$ where $m_{i,j}\geq m_{i,j'}$ for each $0\leq j'\leq q-1$.

Based on the set of output words $Y$, the \emph{majority algorithm} with some threshold $\tau\geq N/2$ outputs the word $\text{maj}_{\tau}(Y) = (z_1, z_2, \ldots, z_n) \in \F_q^n$, where
\[
z_i = 
\begin{cases}
	j & \text{ if } M(i) > \tau \text{ and }M(i)=m_{i,j} \text{ for some } j\in[0,q-1], \\
	? & \text{ otherwise.}
\end{cases} 
\]
In other words, for each coordinate of $\bz$, we choose $?$ or the most frequent symbol $j$ ($j\in[1,q]$) if it is common enough.

Below we introduce a decoding algorithm for substitution errors in the $q$-ary Hamming space ($q\geq3$). The algorithm takes $N_{t,e}$ output words as input and gives the transmitted word $\bx$ as output. The algorithm requires some threshold constants.  Let 
$$\tau'_{t,e}=\frac{1}{e+1}V_q(n-e-1,\ell-1)$$ and 
$$\tau_{t,e}=\tau'_{t,e}+\frac{e}{e+1}N_{t,e}.$$ In  Algorithm \ref{algSubs}, the notation $\mathcal{D}_C(\bu)$ means that the decoder of the code $C$ decodes the word $\bu$ and returns a codeword of $C$. We denote by $\com(\mathcal{D}_C)$ the time complexity of decoder $\mathcal{D}_C$. The algorithm and proofs are inspired by \cite{Uusi_Maria_Abu-Sini}.


\begin{algorithm}
	\caption{Decoding substitution errors in $\F_q^n$}\label{algSubs}
	\begin{algorithmic}[1]
	\Require $N_{t,e}$ output words $Y=\{\by_1,\dots,\by_{N_{t,e}}\}$
	\Ensure Transmitted word $\hat{\bc}$ ($=\bx$) 
\State $\bz=\text{maj}_{\tau_{t,e}}(Y)$
\State $S=\{i\mid i\in[1,n],z_i=?\}$
\State $Z=\{\bu\in \F_q^n\mid u_i=z_i \text{ for all } i\not\in S\}$
\For{each $\bu\in Z$} $\hat{\bc}=\mathcal{D}_C(\bu)$. 
\If{$Y\subseteq B_t(\hat{\bc})$} 
\State\Return $\hat{\bc}$
\EndIf
\EndFor			\end{algorithmic}
	\end{algorithm}

Let $i\in[1,n]$, 
$e_i=|\{\by\in Y\mid y_i\neq x_i\}|$ and $ce_i=|\max_{a\in\F_q}\{\by\in Y\mid y_i\neq x_i\text{ and }y_i=a\}|$ where $\bx$ is the transmitted word. In other words, 
 $e_i$ is the number of output words in $Y$ for which an error has occurred in the $i$th bit and $ce_i$ is the number of output words in $Y$ for which the most common error has occurred in the $i$th bit (there are $q-1$ different error possibilities for each coordinate). Moreover, an error occurs in $z_i$ if $ce_i>\tau_{t,e}$ and having the symbol $?$ at the $i$th position requires that $N_{t,e}-\tau_{t,e}\leq e_i$. Indeed, otherwise $N_{t,e}-e_i>\tau_{t,e}$. Notice that  we do not have a symbol $?$ at the $i$th position even though $N_{t,e}-\tau_{t,e}\leq e_i$ if $ce_i>\tau_{t,e}$.
  
  In the following lemma, we give an upper bound for the number coordinates in which $z_i$ is interpreted as an incorrect symbol (other than $?$).

\begin{lemma}\label{LemAlgErrors}
There are at most $e$ errors in the word $\bz$ in Step $1$.
\end{lemma}
\begin{proof}
Let us assume on the contrary that there are at least $e+1$ errors in $\bz$. We may assume, without loss of generality, that those errors occur in the coordinates $[1,e+1]$. If more than $e+1$ errors occur, then we are interested only in the $e+1$ errors in the coordinates $[1,e+1]$. Let us first count the maximum number of output words which can have $e+1$ common errors among those coordinates. In other words, we count the number of output words which have the same erroneous symbols among the $e+1$ first coordinates. Recall that each output word is unique and thus, this number has an upper bound. Indeed, if we have $e+1$ shared/common errors in the $e+1$ first coordinates, then we can have at most $\ell-1$ errors in the other coordinates. Thus, there are at most $V_q(n-e-1,\ell-1)$ such words.

Let us approximate $ce_i$ for $i\in [1,e+1]$. We have $$ce_i>\tau_{t,e}=\tau'_{t,e}+\frac{e}{e+1}N_{t,e}.$$ Therefore, 
\begin{equation}\label{LBSubsAlgErr}
\sum_{i=1}^{e+1}ce_i>(e+1)\tau'_{t,e}+eN_{t,e}.
\end{equation}

Let us next consider an upper bound for $ \sum_{i=1}^{e+1}ce_i$. As we have seen, there are at most $V_q(n-e-1,\ell-1)=(e+1)\tau'_{t,e}$ output words which can have shared errors in all of the $e+1$ first coordinates. Other $N_{t,e}-V_q(n-e-1,\ell-1)$ output words can have at most $e$ errors in those coordinates. Thus,
\begin{align*}
\sum_{i=1}^{e+1}ce_i\leq&(e+1)^2\tau'_{t,e}+e(N_{t,e}-(e+1)\tau'_{t,e})\\
=&(e+1)\tau'_{t,e}+eN_{t,e}.
\end{align*}
Hence, we have a contradiction between the upper bound and the lower bound in (\ref{LBSubsAlgErr}) and the claim follows.
\end{proof}

Let us then consider the maximum cardinality of the set $S$ in the algorithm. 
\begin{lemma}\label{LemAlg?}
We have $|S|< t(e+2)$ when $n$ is large enough.
\end{lemma}
\begin{proof}
Let us first give a useful approximation for $N_{t,e}$. Recall that $e$, $\ell$ and $q$  are constants with respect to $n$. Hence, we have \begin{align*}
N_{t,e}=&\sum_{i=0}^{\ell-1}\binom{n-2e-1}{i}(q-1)^i\sum_{k=e+i-(\ell-1)}^{t-i}\sum_{j=e+i-(\ell-1)}^{t-i}\binom{2e+1}{k}\binom{2e+1-k}{j}(q-2)^{2e+1-k-j}+1\\
=&\binom{n-2e-1}{\ell-1}(q-1)^{\ell-1}\sum_{k=e}^{e+1}\sum_{j=e}^{e+1}\binom{2e+1}{k}\binom{2e+1-k}{j}(q-2)^{2e+1-k-j}+\Theta(n^{\ell-2})\\
=&\binom{n-2e-1}{\ell-1}(q-1)^{\ell-1}\left(\binom{2e+1}{e}((e+1)(q-2)+2)\right)+\Theta(n^{\ell-2}).
\end{align*}
In the following, we consider the term of the previous sum with $n^{\ell-1}$. When $n$ is large enough, we have \begin{align}
&\binom{n-2e-1}{\ell-1}(q-1)^{\ell-1}\left(\binom{2e+1}{e}((e+1)(q-2)+2)\right)\nonumber\\
\geq&(q-1)^{\ell-1}(e+2)\binom{n-2e-1}{\ell-1}\binom{2e+1}{e}+(q-1)^{\ell-1}\binom{n-2e-1}{\ell-1}\binom{2e+1}{e}\nonumber\\
\geq& (q-1)^{\ell-1}(e+2)\binom{n-2e-1}{\ell-1}\binom{2e+1}{e}+(e+2)V_q(n-e-1,\ell-2)\label{IneqStar1}\\
=&(e+2)\left(\frac{\prod_{i=1}^{e}n-e-t+i}{\prod_{j=1}^{e} n-2e-1+j}\binom{2e+1}{e}(q-1)^{\ell-1}\binom{n-e-1}{\ell-1}+V_q(n-e-1,\ell-2)\right)\nonumber\\
\geq&(e+2)\left((q-1)^{\ell-1}\binom{n-e-1}{\ell-1}+V_q(n-e-1,\ell-2)\right)\label{IneqStar2}\\
=& (e+2)V_q(n-e-1,\ell-1).\nonumber
\end{align} Above, Inequality (\ref{IneqStar1}) follows with the observations $\binom{n-2e-1}{\ell-1}\in \Theta(n^{\ell-1})$ and $V_q(n-e-1,\ell-2)\in \Theta(n^{\ell-2})$. In Inequality (\ref{IneqStar2}), we have $\frac{\prod_{i=1}^{e}n-e-t+i}{\prod_{j=1}^{e} n-2e-1+j}\binom{2e+1}{e}\geq \left( \frac{n-e-t+1}{n-e-1}\right)^{e}\binom{2e+1}{e}\geq1$, when $n$ is large and $e\geq1$. Thus, for large enough $n$, we may use $N_{t,e}/(e+2)> V_q(n-e-1,\ell-1)$ and $\tau'_{t,e}< \frac{1}{(e+1)(e+2)}N_{t,e}$. 


We have $N_{t,e}$ channels, in each of which at most $t$ errors occurs. Thus, $\sum_{i=1}^ne_i\leq t N_{t,e}$. Moreover, for each element in $S$, we require at least $N_{t,e}-\tau_{t,e}$ errors. 
Thus, $$|S|\leq \frac{tN_{t,e}}{N_{t,e}-\tau_{t,e}}=t\frac{N_{t,e}}{\frac{1}{e+1}N_{t,e}-\tau'_{t,e}}<t\frac{N_{t,e}}{\frac{1}{e+1}(\frac{e+1}{e+2}N_{t,e})}= t(e+2).$$
\end{proof}

Finally, we conclude  in the next theorem that the algorithm works as intended with complexity $\Theta(nN_{t,e})=\Theta(n^\ell)$ (provided that $\com(\mathcal{D}_C)\in O(n^\ell))$. Indeed, we have $N_{t,e}\in \Theta(n^{\ell-1})$, see Theorem \ref{LevL=1}. Notice that $\Theta(n^\ell)$ is the optimal complexity in the sense that it takes $\Theta(nN_{t,e})$ time to read all the input words. 
\begin{theorem}\label{AlgSubsWorks}
The complexity of Algorithm \ref{algSubs} is $\Theta(n^{\ell}+\com(\mathcal{D}_C))$ and it outputs the transmitted codeword $\bx$ when it has input of $N_{t,e}$ words in $Y$.
\end{theorem}
\begin{proof}
By Lemma \ref{LemAlgErrors}, there are at most $e$ errors in $\bz$. Moreover, by Lemma \ref{LemAlg?}, there are at most $t(e+2)$ symbols ? in $\bz$. Thus, there are at most $q^{t(e+2)}$ different possible choices for $\hat{\bc}$ in Step 4, when we go through all the possible combinations of symbols that might occur in the coordinates with $?$. Thus, at least one of these words $\bu$ in Step $4$ is identical in these particular coordinates to the transmitted word $\bx$. Moreover, since there are at most $e$ errors in this word, the decoder $\mathcal{D}_C$ correctly decodes this word as $\bx$. Therefore, the transmitted word $\bx$ is among the codewords $\hat{\bc}$. Finally, $\bx$ is the only word which we can output, since we have $N_{t,e}$ channels and there is only one codeword in $\bigcap_{\by\in Y}B_t(\by)$ by Theorem \ref{LevL=1}. Thus, the algorithm works.

We observe that Step $1$ has complexity $\Theta(Nn)$. Steps $2$ and $3$ have complexity $\Theta(n)$. Finally, Steps 4 -- 8 have  complexity  $q^{t(e+2)}\com(\mathcal{D}_C)$ (here  $q^{t(e+2)}$ is constant) plus complexity of $q^{t(e+2)}\Theta(Nn)$ since it takes $\Theta(Nn)$ time for each candidate $\hat{\bc}$ to test whether $Y\subseteq B_t(\hat{\bc})$. Thus, the claim holds.
\end{proof}

\medskip

In the following, we introduce  a list-decoding algorithm for substitution errors  in binary Hamming spaces.
In \cite{junnila2022levenshtein}, the authors have studied substitution errors in the \emph{binary} Hamming space together with list-error-correcting codes. In what follows, we first recall some useful results for the list size $\LL$ previously presented in \cite{junnila2022levenshtein}. Then, we proceed by providing a new decoding algorithm of optimal time complexity list-error-correcting codes.  In this section, we assume that $C$ is a list error-correcting code with $|B_{e+a}(\bu)\cap C|\le M$ for some constants $M$ and  $a\in [0,\ell-1]$.

The first theorem gives a general upper bound. 

\begin{theorem}[Theorem 21, \cite{junnila2022levenshtein}] \label{tldc}
	Let $N\ge V_2(n,\ell-a-1)+1$ where  $0\leq a\leq \ell-1$. Let $C$ be an $e$-error-correcting code such that $|B_{e+a}(\bu)\cap C|\le M$ for every $\bu\in \F_2^n$. Consequently,
	$$\LL\le 2^{\ell-a}M.$$ 
\end{theorem}

The second useful theorem gives an improved upper bound but requires that $n$ is rather large.

\begin{theorem}[Theorem 26, \cite{junnila2022levenshtein}]\label{list(t+2)M}
	Let $N\geq V_2(n,\ell-a-1)+1$, $n\geq (\ell-a-1)^22^b+\ell-a-2$, $\ell-1\geq a\geq1$ and $b=\left\lceil(2e+2a+2)^{\mathbbm{e}\cdot(e+a+1)!}\right\rceil$. Moreover, let $C$ be such an $e$-error-correcting code that there are at most $M$ codewords in any $(e+a)$-radius ball. Then $$\LL\leq \max\{(t+1)M,b/(2e+2a+2)\}.$$
\end{theorem}

The decoding algorithm we are using is quite similar to Algorithm \ref{algSubs} and the algorithm in \cite{Uusi_Maria_Abu-Sini}. One of the main differences is that we introduce a new variable $a\in [0,\ell-1]$ which affects our choices for threshold constant $\tau$. Let 
$$\tau'_{t,e,a}=\frac{1}{e+a+1}V_2(n-e-1-a,\ell-1-a)$$ 
and 
$$\tau_{t,e,a}=\tau'_{t,e,a}+\frac{e+a}{e+a+1}N_{t,e,a}.$$ 
Moreover, let us have $N_{t,e,a}=\left\lceil(1+\epsilon)V_2(n-e-1-a,\ell-1-a)\right\rceil+1$ for some constant $\epsilon>0$. 
Let $\mathcal{D}_C$ be a decoder for $C$ such that, given a word $\bu\in \F_2^n$, the decoder  outputs a list $\mathcal{D}_C(\bu)$ of codewords such that $C\cap B_{e+a}(\bu)\subseteq \mathcal{D}_C(\bu)$ and $|\mathcal{D}_C(\bu)|\leq M$.

\begin{algorithm}
	\caption{List decoding substitution errors in $\F_2^n$}\label{algListSubs}
	\begin{algorithmic}[1]
		\Require  $N_{t,e,a}$ output words $Y=\{\by_1,\dots,\by_{N_{n,t,e,a}}\}$
		\Ensure List of possible transmitted words $T(Y)$ with $\bx\in T(Y)$ 
		\State $\bz=\text{maj}_{\tau_{t,e,a}}(Y)$
		\State $S=\{i\mid i\in[1,n],z_i=?\}$
		\State $Z=\{\bu\in \F_2^n\mid u_i=z_i \text{ for all } i\not\in S\}$
		\For{each $\bu\in Z$} calculate $\mathcal{D}_C(\bu)$
		\For{each $\hat{\bc}\in\mathcal{D}_C(\bu)$}  
		\If{$Y\subseteq B_t(\hat{\bc})$} 
		\State Add $\hat{\bc}$ to $T(Y)$
		\EndIf
		\EndFor
		\EndFor	
		\State\Return $T(Y)$		\end{algorithmic}
\end{algorithm}

Observe that since we are now considering binary Hamming space, an error occurs in the $i$th coordinate if and only if $e_i>\tau_{t,e,a}$. Moreover, we have the symbol ? in the $i$th coordinate if $N_{t,e,a}-\tau_{t,e,a}\leq e_i\leq\tau_{t,e,a}$. Let us first show that  we have at most $e+a$ errors in the majority word $\bz$.

\begin{lemma}\label{LemAlgErrorsList}
	There are at most $e+a$ errors in the word $\bz$ in Step $1$.
\end{lemma}
\begin{proof}
	As the proof is quite similar to the proof of Lemma \ref{LemAlgErrors}, we postpone the proof to Appendix.
\end{proof}

Let us then consider the maximum cardinality of the set $S$ in the algorithm. 
\begin{lemma}\label{LemAlg?List}
	We have $|S|< \frac{t(e+a+1)(1+\epsilon)}{\epsilon}$.
\end{lemma}
\begin{proof}
	Let $N_{t,e,a}=\lceil(1+\epsilon)V_2(n-e-a-1,\ell-a-1)\rceil+1=(1+\epsilon)V_2(n-e-a-1,\ell-a-1)+1+c$ where $0\leq c<1$ is a constant.
	We have $N_{t,e,a}=\lceil(1+\epsilon)V_2(n-e-a-1,\ell-a-1)\rceil+1=(1+\epsilon)(e+a+1)\tau'_{t,e,a}+1+c$ channels, in each of which at most $t$ errors occurs. Thus, $\sum_{i=1}^ne_i\leq t N_{t,e,a}$. Moreover, for each element in $S$, we require at least $N_{t,e,a}-\tau_{t,e,a}$ errors.

		%
		%
		%

		Thus, \begin{align*}
			|S|\leq& \frac{tN_{t,e,a}}{N_{t,e,a}-\tau_{t,e,a}}\\
			=&t\frac{N_{t,e,a}}{\frac{1}{e+a+1}N_{t,e,a}-\frac{1}{e+a+1}V_2(n-e-1-a,\ell-1-a)}\\
			=&t\frac{(e+a+1)((1+\epsilon)V_2(n-e-a-1,\ell-a-1)+1+c)}{(1+\epsilon)V_2(n-e-a-1,\ell-a-1)-V_2(n-e-a-1,\ell-a-1)+1+c}\\
			=&\frac{t(e+a+1)(1+\epsilon)}{\epsilon}\cdot\frac{V_2(n-e-a-1,\ell-a-1)+\frac{1+c}{1+\epsilon}}{V_2(n-e-a-1,\ell-a-1)+\frac{1+c}{\epsilon}}\\
			<& \frac{t(e+a+1)(1+\epsilon)}{\epsilon}.
		\end{align*}
	\end{proof}

	Finally, we conclude that the algorithm works as intended with optimal (assuming that we read all the output words) complexity $\Theta(Nn)$ when $\epsilon$ and $M$ are constants, and the complexity of the list-decoder $\com(\mathcal{D}_C)$ belongs to $O(Nn)$. 
	
	\begin{theorem}\label{ALGListOptComp}
		Algorithm \ref{algListSubs} outputs a list $T_D(Y)$, with $\bx\in T_D(Y)$ and $|T_D(Y)|\leq \LL$, 
		and its (optimal) complexity is  $\Theta(Nn+\com(\mathcal{D}_C))$ when the input is $N_{t,e,a}$ words in $Y$.
	\end{theorem}
	\begin{proof}
	As the proof is quite similar to the proof of Theorem \ref{AlgSubsWorks}, we postpone the proof to Appendix.
	\end{proof}

	In  Theorem \ref{ALGListOptComp}, we say that $|T_D(Y)|\leq \LL$. 	
	Some upper bounds for $\LL$ have been presented in  Theorems \ref{tldc} and \ref{list(t+2)M}. These Theorems require that $N\geq V_2(n,\ell-a-1)+1$. Observe that when $n\geq \frac{(t-1)(1+\epsilon)^{1/(e+a+1)}}{(1+\epsilon)^{1/(e+a+1)}-1}$, we have $(1+\epsilon)V_2(n-e-a-1,\ell-a-1)\geq V_2(n,\ell-a-1)$. Indeed, $V_2(n-e-a-1,\ell-a-1)\geq (\frac{n-t+1}{n})^{e+a+1}V_2(n,\ell-a-1)$, since for any integer $b\in[0,\ell-a-1]$ we have
	\begin{align*}
	\binom{n-e-a-1}{\ell-a-1-b}=&\frac{\prod_{i=1}^{e+a+1}n-t+b+i}{\prod_{j=1}^{e+a+1} n-e-a-1+j}\binom{n}{\ell-a-1-b}\\
	\geq&\left( \frac{n-t+1}{n}\right)^{e+a+1}\binom{n}{\ell-a-1-b}
	\end{align*}
	and $(1+\epsilon)(\frac{n-t+1}{n})^{e+a+1}\geq1$ when  $(1+\epsilon)^{1/(e+a+1)}\geq \frac{n}{n-t+1}$. Moreover, this last condition is satisfied when $n\geq \frac{(t-1)(1+\epsilon)^{1/(e+a+1)}}{(1+\epsilon)^{1/(e+a+1)}-1}>t-1$. Hence, when this requirement is fulfilled, we may use Theorems \ref{tldc} and \ref{list(t+2)M}. Overall, the smaller the constant $\epsilon$ is, the less channels we require. However, we have a trade-off; smaller $\epsilon$ requires larger $n$ and the complexity of the algorithm will increase (slightly) as we will see in  Theorem \ref{ALGListOptComp} based on Lemma \ref{LemAlg?List}.

	In particular, Theorems \ref{tldc} and \ref{list(t+2)M} offer following bounds for $\LL$. If $a\in[1,\ell-1]$, $|B_{e+a}(\bu)\cap C|\leq M$ for every $\bu\in \F_2^n$, $b=\lceil(2e+2a+2)^{\mathbbm{e}\cdot(e+a+1)!}\rceil$ and $ n\geq (\ell-a-1)^22^b+\ell-a-2$, then $\LL\leq \min\{2^{\ell-a}M,\max\{(t+1)M,b/(2e+2a+2)\}\}$ and if $n<(\ell-a-1)^22^b+\ell-a-2$, then $\LL\leq 2^{\ell-a}M$. Moreover, if $a=0$, then $\LL\leq 2^\ell$ (in this case $M=1$ since $C$ is an $e$-error-correcting code).

	\begin{remark}\label{RemarkL=2}
		Observe that when $a=0$ (and $M=1$), by choosing a suitable $\epsilon$, we can find a number of channels which gives an efficient decoding algorithm for the case $\LL\leq2$ by using  \cite[Theorem 6]{yaakobi2018uncertainty} or \cite[Corollary 16]{junnila2022levenshtein}. Indeed, now $N_{t,e,0}=\left\lceil(1+\epsilon)V_2(n-e-1,\ell-1)\right\rceil+1\geq V_2(n,\ell-1)+1$ when $n\geq\frac{(t-1)(1+\varepsilon)^{1/(e+1)}}{(1+\varepsilon)^{1/(e+1)}-1}$ and hence, the number of channels is large enough for $\LL\leq2$ when $n$ is large enough.
	\end{remark}


\section{Probabilities with large \texorpdfstring{$q$}{q} for various error types}\label{SecLargeq}

Next, we consider the likelihood for $|T(Y)|=1$ when $q$ is large, other parameters are constants and only single error type occurs. 
In particular, we show that insertion, substitution and deletion errors behave differently while deletion and erasure errors have similar behaviours. In this section, we work under the assumption that every output word is equally probable and that we may obtain the same output word from multiple channels. In other words, given a transmitted word $\bx$,  the probability that we obtain a given ordered multiset $Y$ of $N$ output words  is $1/|B_t(\bx)|^N$ where $B_t(\bx)$ is the $t$-radius ball centred at $\bx$ with suitable error-metric. Note that a case in which each channel having a unique set of errors (instead of unique output word) has been considered  in \cite{junnila2023levenshtein, junnila2024unique}. These approaches differ in the cases of insertion and deletion errors.

In the following theorem, we first consider \textit{insertion} errors together with large $q$ from probabilistic viewpoint. The result can be  somewhat surprising as in \cite[Equation (51)]{levenshtein2001efficient}, Levenshtein has shown that increasing $q$ implies that we require more channels to have $\LL=1$, that is, $|T(Y)|=1$ for all $Y$. 

\begin{theorem}\label{InsProbqInf}
	Let $N\geq2$ and $C\subseteq\F_q^n$. Assume that a word $\bx$ is transmitted through $N$ channels and a multiset of output words $Y$ is obtained. Moreover, at most $t$ insertion errors occur in each channel. Then the probability that $|T(Y)|=1$ approaches $1$ as $q$ grows and we have  a probabilistic decoding algorithm with  complexity $O(q)$ which never returns an incorrect result.
\end{theorem}
\begin{proof}
	Let $\by_1$ and $\by_2$ be any (possibly $\by_1=\by_2$) output words of $Y$. Let their lengths be $n+t_1$ and $n+t_2$, respectively where $t_1,t_2\leq t$. Thus, there are at most $2t$ new symbols in $\by_1$ and $\by_2$ which are not present in the transmitted word $\bx$. Moreover,  there are at most $n$ different symbols in $\bx$. 
	Recall that we assumed every output word to have the same probability. We note that when the insertion error inserts a symbol which already appears in an arbitrary word $\bw$, we may arrive to the same output word with multiple different insertion errors. For example, if $\bw=0$, then inserting 0  as the first or the second symbol, leads to the same word $00$ while if we had inserted symbol $1$, then we could obtain words $10$ and $01$.  
In general, inserting symbols not present in $\bx$ can result in more unique output words than inserting symbols which appear in $\bx$.	
	Thus, there are more combinations of insertions leading to output words with inserted symbol not present in $\bx$ than there are combinations leading to output words with inserted symbol being in $\bx$. Hence, the probability that the first inserted symbol is not in $\bx$, is at least $(q-n)/q$ and for the $j$th inserted symbol the probability for not appearing in $\bx$ or being already inserted is at least $(q-n-(j-1))/q$. 	Since $\by_1$ has at most $t$ symbols not in $\bx$,	the probability that among the $t_2$ symbols inserted to $\bx$ to form $\by_2$, we are only using symbols which are not present in the word $\bx$ or in $\by_1$ and the $t_2$ inserted symbols all differ from each other, is at least 
	$$\prod_{i=0}^{t_2-1}\frac{q-n-t-i}{q}\geq\prod_{i=0}^{t-1}\frac{q-n-t-i}{q}\geq \left(1-\frac{n+2t-1}{q}\right)^{t}\overset{q\to\infty}{\to}1.$$

	Furthermore, any symbol which appears in $\bx$, appears at least twice in the words $\by_1$ and $\by_2$ (at least once in both of these words). Thus, by removing any symbol from $\by_2$ which does not appear in $\by_1$, we are very likely to obtain the transmitted word $\bx$ as $q$ tends to infinity.  Finally, notice that we do only simple operations on $\by_2$. Thus, the time complexity of the decoding algorithm is $O(q)$. Moreover, we may verify whether the decoding algorithm gives a correct result since we know that the transmitted word has length $n$ and we can check whether $\by_2$ has length $n$ after we have deleted the symbols. Indeed, if after the deletions $\by_2$ has length $n$, then it forms the transmitted word $\bx$ since we have deleted only symbols which  have not belonged to $\bx$.\end{proof}





In the following proposition, we show that \textit{substitution} errors do not behave similarly as insertion errors for any set of parameters as $q$ grows.

\begin{proposition}\label{SubsProbqInf2}
	Let $N\geq1$ and $C=\F_q^n$. Assume that there occur at most $t>0$ substitution errors in each channel. Then the probability for $|T(Y)|=1$ does not approach $1$ as $q$ grows for any transmitted word and set of parameters $n,N$ and $t$.
\end{proposition}
\begin{proof}
	We first show that the probability that exactly $t$ errors occur in a channel tends to $1$ as $q$ increases. Notice that exactly $i$ errors may occur in $(q-1)^i\binom{n}{i}$ ways. Hence, the probability is \begin{align*}
		&\frac{(q-1)^t\binom{n}{t}}{\sum_{i=0}^t(q-1)^i\binom{n}{i}}\\
		=&1-\frac{\sum_{i=0}^{t-1}(q-1)^i\binom{n}{i}}{\sum_{i=0}^{t}(q-1)^i\binom{n}{i}}\\
		\geq &1-\frac{\sum_{i=0}^{t-1}(q-1)^i\binom{n}{i}}{(q-1)^{t}}\overset{q\to\infty}{\to}1.
	\end{align*} Moreover, when exactly $t$ substitution errors occur in each channel, the probability that there occurs an error in each channel in the $i$th coordinate, for some $i\in [1,n]$, depends only on the parameters $n$, $N$ and $t$. Indeed, for a single channel, that probability is $(q-1)\cdot\binom{n-1}{t-1}(q-1)^{t-1}/\left(\binom{n}{t}(q-1)^{t}\right)=\binom{n-1}{t-1}/\binom{n}{t}$. Hence, the probability that each channel has a substitution error at the $i$th coordinate is  $P=\left(\binom{n-1}{t-1}/\binom{n}{t}\right)^N$. Furthermore, the probability $P$ is a positive constant since $n$, $N$ and $t$ are constants. Since $C = \F_q^n$ and assuming that a substitution error occurs in the $i$th coordinate of each output word of $Y$, 
	we have $|T(Y)|>1$.  Indeed, for example, if a substitution error occurs in each channel in the first symbol, then $\bx,\bx'\in T(Y)$ where $\bx'$ is such that $x'_1=x_1+1$ and $x'_i=x_i$ for each $i\in[2,n]$.
Thus, the probability that $|T(Y)|=1$ approaches at most $1-P$, instead of $1$. Thus, the claim follows.
\end{proof}

We can also see that in the case of substitution errors, the increase of $q$ may affect on the probability that $|T(Y)|=1$. Consider for example the case with $n=2, t=1, N=3, \bx=00$. The number of possible output words is $1+2(q-1)$. Moreover, we have $|T(Y)|=1$ if and only if each output word is unique and at least one substitution error occurs in the first coordinate for some output word $\by_1$ and in the second coordinate in another output word $\by_2$.


Moreover, there are $2q(q-1)(q-2)$ possible ordered combinations for output words $\by_1,\by_2,\by_3$ such that they are unique and substitutions resulting to them occur either all in the first coordinate or all in the second coordinate.
 Beyond these ordered output word sets, there are also $(1+2(q-1))^3 - (1+2(q-1))(2(q-1))(2(q-1)-1)$ possible
  output word sequences with at least two identical
   output words. Therefore, the probability that we may deduce $\bx$ is \begin{align*}
P'=&\frac{(1+2(q-1))^3-((1+2(q-1))^3 - (1+2(q-1))(2(q-1))(2(q-1)-1))-2q(q-1)(q-2)}{(1+2(q-1))^3}\\
%
=&\frac{ 6q^3-18q^2+18q-6}{8q^3-12q^2+6q-1}=\frac{ 6(q-1)^3}{(2q-1)^3}.
\end{align*}
We notice in particular, that $P'$ is not a constant on $q$ and therefore, the choice for $q$ affects on the probability whether we may deduce $\bx$ from $Y$. Notice that the fact that probability $P'$ depends on $q$, does not contradict the proof of Proposition \ref{SubsProbqInf2} since we only claimed that the probability $P$ has an upper bound with value less than $1$ which is independent of $q$.

Finally, let us consider \textit{deletion} and \textit{erasure} errors with $q>n\geq2$. In the following proposition, for clarity, we avoid writing $\bx=\bx'$ for two identical words when they are considered over different alphabets. For example, for $\bx=00\in \F^2_4$ and $\bx'=00\in \F^2_6$ we say that the words are \emph{identical} rather than write $\bx=\bx'$. Similarly, two sets over different alphabets are called \emph{identical} if their words are identical.

\begin{proposition}\label{DelEraProbqInf}
	Let $N,n$ and $t$ be constant. 
	Further, let $q'>q>n$ and $C=\F_q^n$. Let $\bx\in \F_q^n$ and $\bx'\in\F_{q'}^n$ be identical to $\bx$. Then, sets $B_t^d(\bx)$ (resp. $B_t^e(\bx)$) and $B_t^d(\bx')$ (resp. $B_t^e(\bx')$) are identical and we have 
$$P(T(Y)=\{\bx\})=P(T(Y')=\{\bx'\})$$	
	for the probabilities over all possible output word multisets $Y$ and $Y'$ obtained from $\bx$ and $\bx'$ with  deletion  (resp. erasure) errors.
%
	\end{proposition}
\begin{proof}
We prove this proposition for deletion errors. The proof for erasures is similar. First we show that deletion balls $B_t^d(\bx)$ and $B_t^d(\bx')$ are identical. We may observe that if a word $\by$ can be obtained from $\bx$ with at most $t$ deletions, then it can also be obtained from $\bx'$ with at most $t$ deletions and vice versa. Thus, the two balls are identical. Therefore, the probability for a particular ordered  multiset $Y$ or $Y'$ of output words to occur is $1/|B_t^d(\bx)|^N$ where $Y$ and $Y'$ are two identical output word sets one of which is obtained from $\bx$ and one from $\bx'$.

As $q' > q > n$, we may assume without loss of generality  that the symbols of the identical words $\bx$ and $\bx'$ belong to the set $\{0, 1, 2, \ldots, n-1\}$. In the following, we show that if $Y$ and $Y'$ are identical, then $T(Y)=\{\bx\}$ if and only if $T(Y')=\{\bx'\}$.
Assume first that $T(Y)\neq \{\bx\}$. In particular, let $\bc\neq\bx$ be such that $\bc\in T(Y)$. Notice that in this case we have $\bc'$, a word which is identical to $\bc$, in the set $ T(Y')$ and hence $T(Y')\neq \{\bx'\}$. Assume then that $T(Y)=\{\bx\}$ and $T(Y')\neq \{\bx'\}$. We have $\bx'\in T(Y')$. Hence, we may assume that $\bx'\neq \bc'\in T(Y')$. Notice that if there exists $\bc\in \F^n_q$ which is identical to $\bc'$, then $\bc\in T(Y)$, a contradiction. Hence, $\bc'$ has symbols in $\{q,q+1,\dots,q'-1\}$. Consider next a word $\bc''\in \F^n_q$ which is identical to $\bc'$ except that each symbol in $\{q,q+1,\dots,q'-1\}$ is converted to $n$ (note that since $n<q$, we have $n\in \F_q$). Since $\bc'\in T(Y')$, we have $\bc''\in T(Y)$. Indeed, since $Y$ and $Y'$ are identical, no word in $Y'$ contains a symbol in $\{q,q+1,\dots,q'-1\}$. Thus, if the transmitted word had contained any of them, then we would have deleted each of them. Furthermore, in this case the resulting output word is unaffected by whether the deleted symbol was $n$ or some $n'>n$. Hence, $\bc''\in T(Y)$. Recall that $n$ does not belong to the presentation of $\bx$. Thus, $\bc''\neq\bx$ and $T(Y)\neq\{\bx\}$, a contradiction.

We have now shown, that if $Y$ and $Y'$ are identical, then $T(Y)=\{\bx\}$ if and only if we have $T(Y')=\{\bx'\}$.
 Furthermore, we have shown that if the probability for obtaining ordered multiset $Y$ from $\bx$ is $P$, then the  probability for obtaining identical ordered multiset $Y'$ from $\bx'$ is also $P$. Now, the claim follows from these two observations.
%
\end{proof}

To conclude, we have observed that when $q$ tends to infinity, deletions, substitutions and insertion errors operate in different ways. In the case of insertion errors, the probability that $|T(Y)|=1$ tends to $1$. For substitution errors,  the probability that $|T(Y)|=1$ is bounded from above by a constant other than $1$. Finally, in the case of deletion (or erasure) errors, the probability that $|T(Y)|=1$, is unaffected by the growth of $q$ when $q>n$. The result for insertion errors is somewhat surprising.

\section{Conclusion}\label{SecConclusion}

In this paper, we have focused on the Levenshtein's reconstruction problem in the case of $q$-ary alphabets with respect to various error types. One of our main objectives has been generalizing known results for binary words to $q$-ary cases:
\begin{itemize}
	\item Previously, in~\cite{ junnila2020levenshtein}, the exact number of channels required to obtain a constant list size $\LL$ (with respect to $n$) has been determined in the binary case. In Section~\ref{SecqlistComb}, we generalized the result for the $q$-ary alphabet (with $q \geq 3$). Moreover, our result gave an insight to the number of words in the intersection of Hamming balls of radius $t$, which is a rather difficult problem in general.
	
	\item For substitution errors, a majority algorithm of optimal complexity for decoding the submitted word has been presented for $e=0$ (\cite{levenshtein2001efficient}), $q = 2$ and $e = 1$ (\cite{yaakobi2018uncertainty}), and $q = 2$ and $e > 1$ (\cite{Uusi_Maria_Abu-Sini}). In Section~\ref{Sec:DecodingQ}, we extended these results to the cases with $q \geq 3$ and $e \geq 1$. Moreover, we presented a list-decoding version of the algorithm for $q = 2$. Notice that here we are restricted to the case with $q = 2$ due to the fact that some underlying results (of \cite{junnila2022levenshtein}) required for the algorithm are only known for the binary words. Hence, a natural direction for future research would be to generalize the results of \cite{junnila2022levenshtein} to larger $q$ and then extend the list-decoding algorithm for general $q$.
\end{itemize}

The theme of $q$-ary alphabets also continues in other sections of the paper:
\begin{itemize}
	\item In Section~\ref{SecErasure}, we studied erasure errors in $q$-ary alphabets (with $q \geq 2$), where the introduction of the symbol $*$ corresponding to the erasure error could be interpreted as switching from a $q$-ary alphabet to one with $q+1$ symbols. Then, in Section~\ref{SecErasureSubstitution}, we combine erasures with substitution errors. For future studies, combining erasures with other error types would be interesting and natural. 
	
	\item In Section~\ref{SecLargeq}, we investigated the behaviour of different error types when the size $q$ of the alphabet grows (other parameters remain constant). We notice that under these circumstances different error types behave differently.
\end{itemize}

\bibliographystyle{abbrv}
\bibliography{TCS}

\newpage

\section*{Appendix: Decoding algorithm for list-error-correcting codes}\label{Sec:DecodingList}

	\noindent \textbf{Lemma \ref{LemAlgErrorsList}.} 
		\emph{	There are at most $e+a$ errors in the word $\bz$ in Step $1$.}
\begin{proof}
	Let us assume on the contrary that there are at least $e+a+1$ errors in $\bz$. We may assume, without loss of generality, that those errors occur in the coordinates $[1,e+a+1]$. If more than $e+a+1$ errors occur, then we are interested only in the $e+a+1$ errors in the coordinates $[1,e+a+1]$. Let us first count the maximum number of output words, which can have $e+a+1$ errors in  those coordinates.  Recall that each output word is unique and thus, this number has an upper bound. Indeed, if we have $e+a+1$  errors in the $e+a+1$ first coordinates, then we can have at most $\ell-a-1$ errors in the other coordinates. Thus, there are at most $V_2(n-e-a-1,\ell-a-1)$ such words.
	
	Let us approximate $e_i$ for $i\in [1,e+a+1]$. We have $$e_i>\tau_{t,e,a}=\tau'_{t,e,a}+\frac{e+a}{e+a+1}N_{t,e,a}.$$ Therefore, 
	\begin{equation}\label{LBSubsAlgErrList}
		\sum_{i=1}^{e+a+1}e_i>(e+a+1)\tau'_{t,e,a}+(e+a)N_{t,e,a}.
	\end{equation}

	Let us next consider an upper bound for $ \sum_{i=1}^{e+a+1}e_i$. As we have seen, there are at most $V_2(n-e-a-1,\ell-a-1)=(e+a+1)\tau'_{t,e,a}$ output words which can have errors in all of the $e+a+1$ first coordinates. Other $N_{t,e,a}-V_2(n-e-a-1,\ell-a-1)$ output words can have at most $e+a$ errors in those coordinates. Thus,
	\begin{align*}
		\sum_{i=1}^{e+a+1}e_i\leq&(e+a+1)^2\tau'_{t,e,a}+(e+a)(N_{t,e,a}-(e+a+1)\tau'_{t,e,a})\\
		=&(e+a+1)\tau'_{t,e,a}+(e+a)N_{t,e,a}.
	\end{align*}
	Hence, we have a contradiction between the upper bound and the lower bound  (\ref{LBSubsAlgErrList}) and the claim follows.
\end{proof}

	\noindent \textbf{Theorem \ref{ALGListOptComp}.} 
		\emph{Algorithm \ref{algListSubs} outputs a list $T_D(Y)$, with $\bx\in T_D(Y)$ and $|T_D(Y)|\leq \LL$, 
		and its (optimal) complexity is  $\Theta(Nn+\com(\mathcal{D}_C))$ when the input is $N_{t,e,a}$ words in $Y$.}
	\begin{proof}
		By Lemma \ref{LemAlgErrorsList}, there are at most $e+a$ errors in $\bz$. Moreover, by Lemma \ref{LemAlg?List}, we have $|S|\leq\frac{t(e+a+1)(1+\epsilon)}{\epsilon}$ where $\epsilon$ is a constant. Thus, we have $|Z|\leq 2^{\lfloor\frac{t(e+a+1)(1+\epsilon)}{\epsilon}\rfloor}$. Now, in Step $4$, we calculate $\mathcal{D}_C(\bu)$ for ${|Z|}$ words $\bu$. Furthermore, the list decoder $\mathcal{D}_C$ outputs a list of at most $M$ codewords $\hat{\bc}$ for each word $\bu$. Now, for each $\hat{\bc}$ we check if $\hat{\bc}\in \bigcap_{\by\in Y}B_t(\by)$ and if it is, then we add it to $T_D(Y)$. Observe that there exists a word $\bu$ such that $d(\bu,\bx)\leq e+a$ and $\bx\in T_D(Y)$ where $\bx$ is the transmitted word. Since we have $N_{t,e,a}\geq V_2(n,\ell-a-1)+1$ channels, cardinality of the set $\bigcap_{\by\in Y}B_t(\by)\cap C$ is bounded from above by Theorems \ref{tldc} and \ref{list(t+2)M}. Thus, the algorithm works.
		
		We observe that Step $1$ has complexity $\Theta(Nn)$. Step $2$ has complexity $\Theta(n)$. Similarly Step $3$ has complexity $\Theta(n)$ since we compute $2^{|S|}$, a constant number, different words of length $n$. Finally, in Steps 4 -- 10, the complexity of calculating codewords $\hat{\bc}\in \mathcal{D}_C(\bu)$ is $\com(\mathcal{D}_C)$, there are at most $M$ codewords in $\com(\mathcal{D}_C)$ and we do it at most $|Z|\leq 2^{\lfloor\frac{t(e+a+1)(1+\epsilon)}{\epsilon}\rfloor}$ times (which is a constant). After that, in Step $5$ checking whether a codeword $\hat{\bc}$ belongs to $\bigcap_{\by\in Y}B_t(\by)\cap C$ has complexity of $\Theta(Nn)$ and we do it at most $|Z|M$ times. Thus, the claim holds.
	\end{proof}

\end{document}